\documentclass[prb,twocolumn,superscriptaddress,amsmath,amssymb]{revtex4-1}
\usepackage{amssymb,amsmath}
\usepackage{color}
\usepackage{graphicx}
\usepackage{dcolumn}
\usepackage{bm}
\usepackage{latexsym,epsfig}
\usepackage{hyperref}

\usepackage{marvosym}
\usepackage{color}
\usepackage[utf8x]{inputenc}

\DeclareUnicodeCharacter{2212}{-}
\begin{document}

\title{A multiconfigurational study of the negatively charged nitrogen-vacancy center in diamond}
\author{Churna Bhandari}
\email{Current Address: The Ames Laboratory, U.S. Department of Energy, Iowa State University, Ames, IA
50011–3020, USA}
\affiliation{Department of Physics, Virginia Tech, Blacksburg, Virginia 24061, USA}
\author{Aleksander L. Wysocki}
\affiliation{Department of Physics, Virginia Tech, Blacksburg, Virginia 24061, USA}
\author{Sophia E. Economou}
\affiliation{Department of Physics, Virginia Tech, Blacksburg, Virginia 24061, USA}
\author{Pratibha Dev}
\affiliation{Department of Physics and Astronomy, Howard University, Washington, DC 20059, USA}
\author{Kyungwha Park}
\email{kyungwha@vt.edu (corresponding author)}
\affiliation{Department of Physics, Virginia Tech, Blacksburg, Virginia 24061, USA}


\begin{abstract}
Deep defects in wide band gap semiconductors have emerged as leading qubit candidates for realizing quantum sensing and information applications. Due
to the spatial localization of the defect states, these deep defects can be considered as artificial atoms/molecules in a solid state matrix. Here we show that unlike single-particle treatments, the multiconfigurational quantum chemistry methods, traditionally reserved for atoms/molecules, accurately describe the many-body characteristics of the electronic states of these defect centers and correctly predict properties that single-particle treatments fail to obtain. We choose the negatively charged nitrogen-vacancy (NV$^-$) center in diamond as the prototype defect to study with these techniques due to its importance for quantum information applications and because its properties are well-known, which makes it an ideal benchmark system. By properly accounting for electron correlations and including spin-orbit coupling and dipolar spin-spin coupling in the quantum chemistry calculations, for the NV$^-$ center in diamond clusters, we are able to: (i) show the correct splitting of the ground (first-excited) spin-triplet state into two levels (four levels), (ii) calculate zero-field splitting values of the ground and excited spin-triplet states, in good agreement with experiment, (iii) determine many-body configurations of the spin-singlet states, and (iv) calculate the energy differences between the ground and exited spin-triplet and spin-singlet states, as well as their ordering, which are also found to be in good agreement with recent experimental data. The numerical procedure we have developed is general and it can screen other color centers whose properties are not well known but promising for applications.

\end{abstract}

%
\date{\today}							

\maketitle

\section{Introduction}


Defects in solid-state systems are naturally formed and can be implanted in a controllable fashion. Individual defects deeply embedded in wide band-gap semiconductors are known to have distinct localized electronic states within the band gap and so they behave similar to atoms or molecules. The prototype of such deep defects is the negatively charged nitrogen-vacancy (NV$^{-}$) center defect in diamond which has been extensively used
for sensing~\cite{MazeNature2008,Balas2008}, for the demonstration of loophole-free Bell inequalities~\cite{Hensen2015}, and for a proof-of-principle of quantum error correction~\cite{Taminiau2014,Waldherr2014}, to name a few among many important experiments and quantum information science applications. Its tremendous success was culminated in recent experimental realization of quantum entanglement between the spins of the NV$^-$ centers over a kilometer range~\cite{Hensen2015}. Single spins of the NV$^-$ center defects were shown to be optically initialized and read out with long spin-lattice relaxation and spin coherence times at room temperature \cite{Robledo2011,Humphreys2018,Jarmola2012,Astner2018,Zhao2012,BarGill2012,Awschalom2018,Bassett2019},
and the electronic spin can be coherently controlled both optically \cite{Yale2013} and via microwave fields \cite{Fuchs2009}.
This prototype defect inspired exploration of other defects, hopefully even more suitable for quantum information science applications, in diamond
and other wide band-gap semiconductors such as the silicon vacancies
and NV center in silicon carbide~\cite{Koehl2011,Widmann2015,Soykal2016,Bockstedte2018,Nagy2019}, the silicon vacancy center in diamond~\cite{Edmonds2008,Gali2013,Pingault2017,Rose2018,Ma2020}, and rare-earth defects in silicon~\cite{Yin2013} or yttrium orthosilicate~\cite{Kornher2020}.


Electronic and magnetic properties of deep defects have been studied using either various levels of {\it ab initio} theory
or phenomenological molecular models based on group theory. In the quest of unexplored, improved defects, {\it ab-initio} theory rather than the molecular model approach can play an essential role in screening candidate defects for quantum information science applications before experimental data are available, because the latter approach requires parameter values such as Coulomb interactions and dipolar spin-spin coupling (SSC) and spin-orbit coupling (SOC) strengths. To that end, the techniques need to be reliable and predict defect properties as accurately as possible.
Although single-particle {\it ab-initio} techniques are extensively used, they have serious limitations for strongly correlated systems,
especially for excited states. For example, density-functional theory (DFT) (as well as the molecular model approach) could not correctly predict
the ordering of the spin-singlet states of the NV$^{-}$ center defect in diamond \cite{GaliPRB08,Lenef1996,Manson2006,RogersNJP08}, which led to a long-standing debate and conflicting results in the community \cite{Manson2006,RogersNJP08,Acosta2010,Batalov2008,Toyli2012}. Recent experimental results resolved this conflict~\cite{Kehayias2013,Goldman2015,Goldman2017}. Furthermore, DFT could not correctly predict either the ordering or the energy difference between the excited spin-triplet and spin-singlet states of the NV$^-$ center defect \cite{GaliPRB08,DelaneyNL10,Goss1996}.
The aforementioned incorrect predictions of DFT highly influence our understanding of optical transitions between the triplet and singlet states referred to as intersystem crossings~\cite{Robledo2011,Manson2006,Batalov2008,Toyli2012}, which are key mechanisms to initialize and readout the spin-polarized states for quantum technology applications.

In order to remedy this limitation, quantum chemistry calculations~\cite{DelaneyNL10,ZyubinJPC07,LinJCP08} were performed for the NV$^{-}$ center
defect in diamond clusters, but the electronic structure of the defect states is not all consistent with experimental data
~\cite{Kehayias2013,Goldman2015,DaviesExpt76}. For example, the ordering of the excited triplet and singlet states and the energy differences
between the singlet states (or the excited triplet and singlet states) does not agree with experiment. As a middle ground,
beyond-DFT {\it ab-initio} results were combined with model Hamiltonians within many-body (perturbation) theory~\cite{Bockstedte2018,Ma2020,Ma2010,ChoiPRB012}, finding agreement with experimental data~\cite{Kehayias2013,Goldman2015,DaviesExpt76,Goldman2017}.
However, this method requires fitting of the {\it ab-initio} results to the model Hamiltonian parameters. More importantly, within this method, accounting for the effects of SOC and SSC is not straightforward. So far, zero-field splitting values induced by SOC and/or SSC have not been
studied within many-body {\it ab-initio} methods.



In this work, we investigate the electronic structure and magnetic properties of an NV$^{-}$ center in diamond by systematically applying
multiconfigurational quantum chemistry methods (beyond DFT) to hydrogen-passivated diamond clusters containing the defect. The critical ingredient
for success in quantum chemistry calculations is to include several defect-localized unoccupied states beyond dangling bond states, which differentiates our case from the previous quantum chemistry calculations~\cite{DelaneyNL10,ZyubinJPC07,LinJCP08}.
By considering full electron correlation among these extra defect states and the dangling bond states, we determine excitation energies
between the ground state and the excited spin-triplet and spin-singlet states as well as the character of the states. Furthermore, using the
quantum chemistry methods, we examine effects of SOC and SSC on the spin-triplet states and identify characteristics of the split levels as well as
the zero-field splitting values. To the best of our knowledge, this work is the first quantum chemistry calculation of the zero-field splitting by SOC and SSC for an NV$^{-}$ center in diamond. Our calculated results of the electronic structure and zero-field splitting are compared to recent
experimental data with which we find agreement ranging from good to excellent.

This paper is structured as follows. In Section II we provide a brief overview of the NV$^-$ center in diamond. In Section III we describe the structures of the clusters that are considered. In Section IV we discuss our procedure of applying the quantum chemistry methods to the diamond clusters, while the technical detail with a flowchart is provided in the Appendix. In Section V we present our results of the energy separations and characteristics of the triplet and singlet states as well as the zero-field splitting in comparison to other theoretical studies and experimental data. In Section VI we provide our conclusion and outlook.

\section{Overview of NV$^-$ Center Defect}

\begin{figure*}[hbt!]
\centering
\hspace*{-0.5cm}(a) \hspace*{3.8cm}(b)  \hspace*{4.7cm} (c)\\
\includegraphics[scale=0.17]{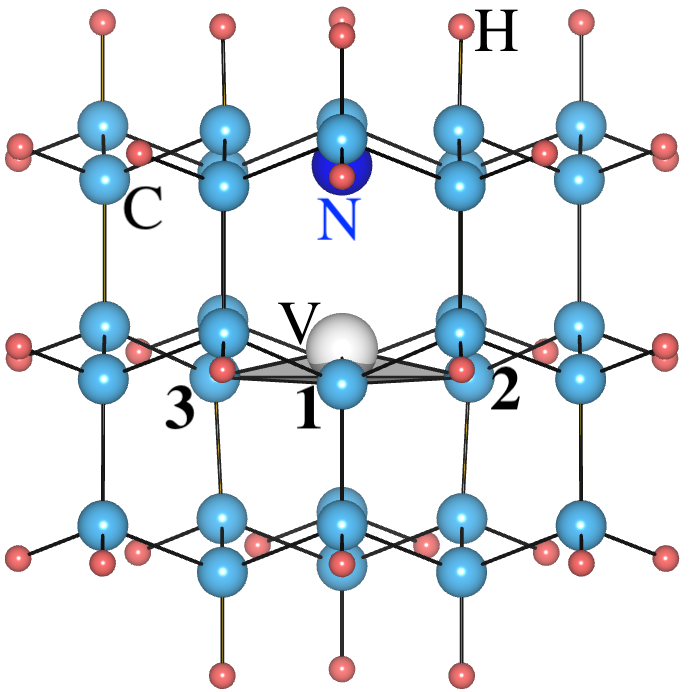}
\includegraphics[scale=0.15]{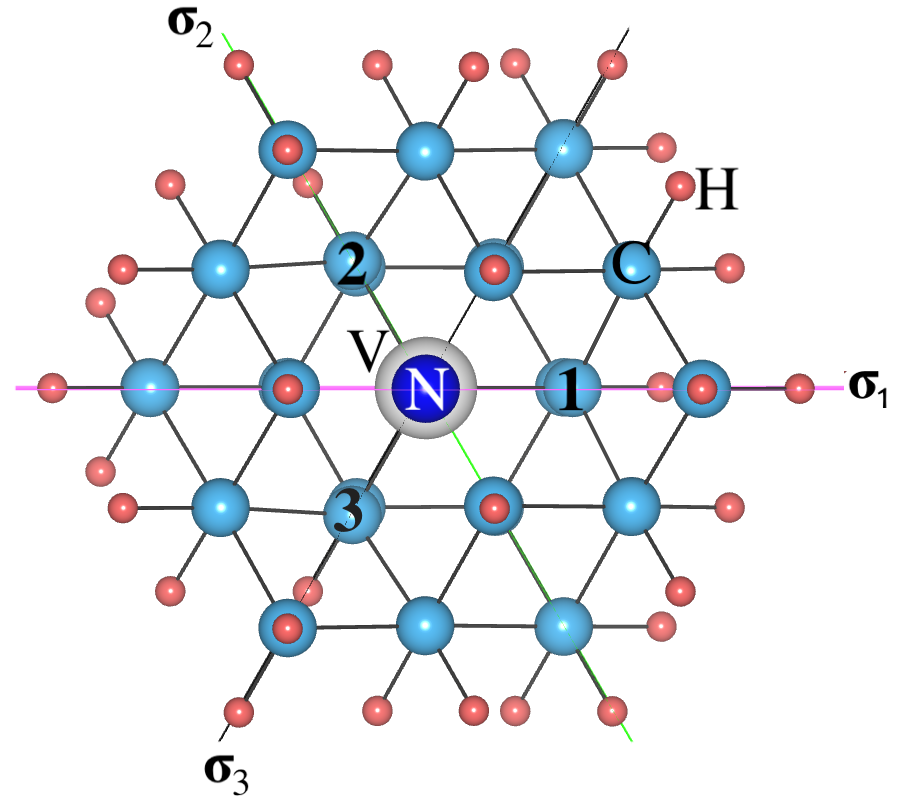}
\includegraphics[scale=0.15]{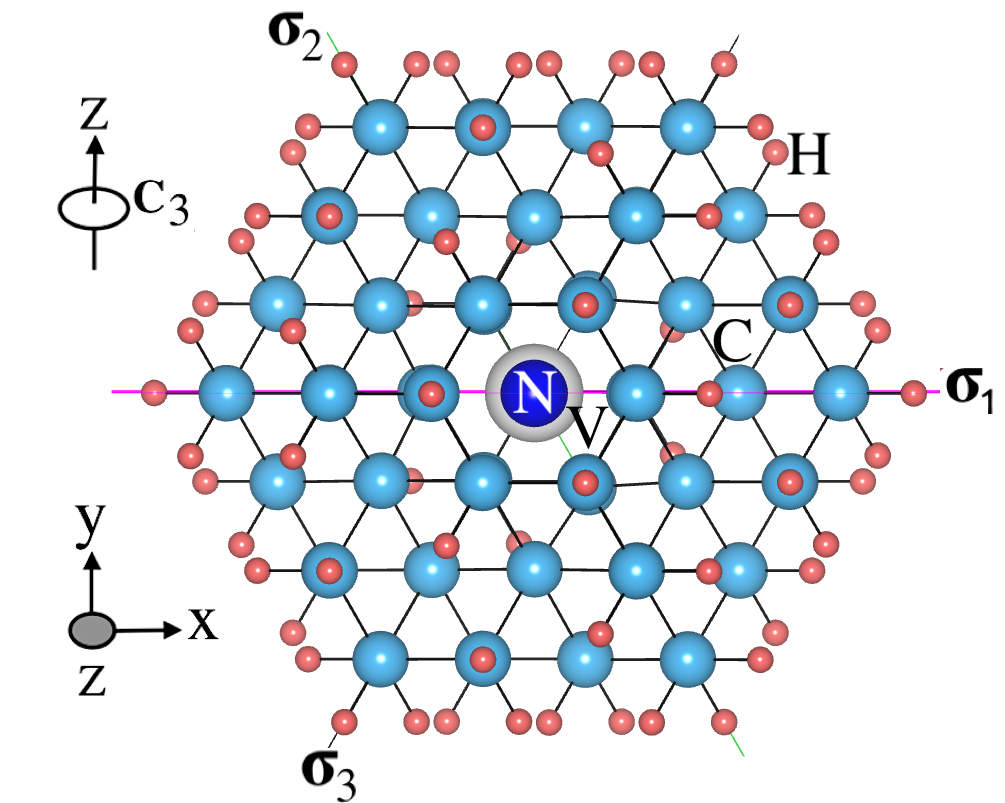}\\
\caption{(a) Side view and (b) top view of the NV$^{-}$ center defect in a 70-atom diamond cluster with $C_{3v}$ symmetry.
(c) Top view of the NV$^-$ center defect in a 162-atom cluster with $C_{3v}$ symmetry. The color scheme is as follows: carbon (cyan), nitrogen
(blue), vacancy (grey), hydrogen (pink). Carbon, nitrogen, vacancy, and hydrogen are denoted by C, N, V, and H, respectively. The rotation axis
of the three-fold symmetry ($C_3$) and the coordinate axes are shown. Here $\sigma_1$ $\sigma_2$, and $\sigma_3$ indicate vertical mirror planes
passing through the carbons nearest to the vacancy with broken dangling bonds (labeled by 1, 2, and 3), the vacancy, and the $z$ axis.}\label{fig1}
\end{figure*}

The deep NV$^{-}$ center defect in diamond consists of a nitrogen atom substituting for carbon and a vacancy at its neighboring carbon site,
as shown in Fig.~\ref{fig1}(a). The axis connecting between the vacancy and nitrogen sites is chosen to be the $z$ axis.
The defect has a $C_{3v}$ point-group symmetry comprising two threefold rotational symmetries ($C_3$) about the $z$ axis and
three vertical mirror planes $\sigma_i$ ($i=1,2,3$) each passing through the nitrogen and nearest carbon atoms in the $xy$-plane
(Fig.~\ref{fig1}).

For an NV$^{-}$ center in diamond, experimental zero-phonon absorption spectra showed that the ground state is a spin-triplet $^3A_2$
state with an excitation energy of 1.945~eV to the first-excited spin-triplet $^3E$~\cite{DaviesExpt76} and that the excitation energy between the
lowest and first-excited spin-singlet states ($^1E$$-$$^1A_1$) is 1.190~eV~\cite{RogersNJP08}. Recent experimental data \cite{Kehayias2013,Goldman2015} showed that the singlet $^1A_1$ state has a higher energy than the singlet $^1E$ state. So far, there have been no direct measurements on the excitation energy of the spin-singlet $^1E$ state relative to the ground $^3A_2$ state. This excitation energy, however, can be deduced from
the experimental energy difference between the $^3E$ and $^1A_1$ states (which is in the range of 0.321 to 0.414~eV~\cite{Goldman2015,Goldman2017}) as
well as from the $^1A_1$$-$$^1E$ energy difference.

\section{Cluster Structures}

To study the NV$^{-}$ center in diamond, we consider two vacancy-centered clusters with hydrogen passivation,
C$_{33}$H$_{36}$N$^{-}$ (70-atom cluster) and C$_{85}$H$_{76}$N$^{-}$ (162-atom cluster), which are created such that they have the correct $C_{3v}$ symmetry. The geometries of the clusters are constructed from
the DFT-optimized, $C_{3v}$-symmetric structure of a 215-atom cubic supercell with an NV$^{-}$ center. The DFT calculation of the relaxation
is performed for the cubic supercell with $4 \times 4 \times 4$ $k$-points within the Perdew-Burke-Ernzerhof (PBE)~\cite{Perdew1996} generalized
gradient approximation using Quantum Espresso~\cite{QE}. Ultrasoft pseudopotentials with scalar relativistic terms and non-local core corrections
are used until the maximum residual force is less than 0.005~eV/\AA.~Figure~\ref{fig1} shows side and top views of the 70-atom cluster and a top
view of the 162-atom cluster where the $z$ axis is along the body-diagonal [111] direction in the cubic supercell. After the geometry optimization,
the $C_{3v}$ point-group symmetry is retained at the NV$^{-}$ center in the supercell. For the DFT-optimized supercell,
the bond length between the nitrogen atom and the carbon atoms nearest to the vacancy is 2.734~\AA,~and the bond lengths between two nearest neighboring carbon atoms closest to the vacancy is 2.676~\AA.~These bond lengths agree well with the corresponding bond lengths reported from
other DFT calculations~\cite{Viktor014}. The shortest distance between the vacancy and carbon (nitrogen) is 1.647~(1.690)~\AA. For the clusters,
the bond length between hydrogen and carbon is set to a standard value, 1.09~\AA,~and no further relaxation is carried out.

\section{Quantum Chemistry Methods}\label{method}

The quantum chemistry calculations are carried out in two steps: (i) complete active space self-consistent field (CASSCF) calculations with state average~\cite{molcas_book}; (ii) inclusion of SOC and SSC. We use both the {\tt OpenMolcas}~\cite{openmolcas} code and the {\tt ORCA}~\cite{orca018,orca012} code. The scalar relativistic effects are included based on the Douglas-Kroll-Hess Hamiltonian using relativistically contracted all-electron correlation-consistent polarized double-zeta basis sets, cc-pVDZ-DK~\cite{ccbasis89,ccDK01}, for all atoms in the clusters.
A schematic flow chart of our computational procedure is shown in Fig.~\ref{fig2}(c).


\subsection{CASSCF calculations}

In the CASSCF formalism~\cite{molcas_book}, a many-body wave function is described as a linear combination of multiple Slater's determinants, each
of which is made of single-electron molecular orbitals. The coefficients of the Slater's determinants are referred to as configuration interaction (CI) expansion coefficients. A CASSCF wave function is partitioned into parts from inactive orbitals with double occupancy, virtual orbitals with zero occupancy, and active orbitals with occupancy between zero and two (i.e., 0, 1, or 2). In a CASSCF calculation, for a given spin multiplicity, any possible electron configurations or correlation within the active orbital space are included, while keeping the occupancies of the inactive and virtual orbitals fixed. However, electron excitation or correlations outside the active space are not included. Both the CI coefficients and the molecular orbitals are optimized through self-consistent calculations. Therefore, the choice of the active orbitals is critical for accurate CASSCF
calculations. It was shown that the accuracy of CASSCF calculations is greatly improved by including extra molecular orbitals beyond frontier orbitals
in the active space~\cite{molcas_book}. CASSCF wave functions are described in terms of spin-free basis states that correspond to all
possible configurations generating the maximum $M_z$ values, where $M_z$ is an eigenvalue of the $S_z$ operator (i.e., the $z$ component of
the total spin $S$). The state-average is a technique to facilitate convergence of the excited-state CASSCF wave functions~\cite{molcas_book}.

In order to determine the number and character of orbitals to be included in the active space, we start with a qualitative analysis of the
electronic structure of an NV$^{-}$ center from a single-electron point of view. The NV$^{-}$ center in diamond has four broken dangling bonds,
as shown in Fig.~\ref{fig1}(a): three dangling bonds of the nearest neighboring carbon atoms to the vacancy ($d_1$, $d_2$, and
$d_3$), and the dangling bond of the nitrogen atom to the vacancy ($d_N$). They form four single-electron molecular orbitals such as $a_{1}^{C}=(d_1+d_2+d_3)/3$, $a_{1}^{N}=d_N$, $e_x=(2d_1-d_2-d_3)$, and $e_y=(d_2-d_3)/\sqrt{2}$~~\cite{Lenef1996,MazeNJP011,DohertyNJP011}.
The first two orbitals transform as a function of the $A_1$ irreducible representation (IRRep), and the other two orbitals transform as functions
of the $E$ IRRep under the C$_{3v}$ point group. It is known that the $a_{1}^{N}$ orbital is deeply buried under the valence band of the diamond
lattice, whereas the other three orbitals are within the band gap~\cite{GaliPRB08,Lenef1996,MazeNJP011,DohertyNJP011}. These three states
are also referred to as in-gap defect states~\cite{Bockstedte2018}. Now let us count the total number of electrons in the system. A carbon vacancy within diamond leaves four electrons in four dangling bonds. One of these carbon-atoms is substituted with a nitrogen-atom that has an extra electron (as compared to a carbon atom). The defect further acquires an additional electron and becomes negatively charged, resulting in a total number of six electrons that fill the defect states in accordance with the Hund's rules. In the spin-triplet ground-state, the {\it nominal} occupancy is as follows: the defect state, $a_{1}^{N}$, which lies in the valence band, is doubly occupied, while the remainder of the four electrons are distributed amongst the in-gap states, with $a_{1}^{C}$ being doubly occupied, and the degenerate orbitals, ($e_x$ $e_y$), being singly occupied.

\begin{figure*}[hbt!]
\centering
\includegraphics[scale=0.325]{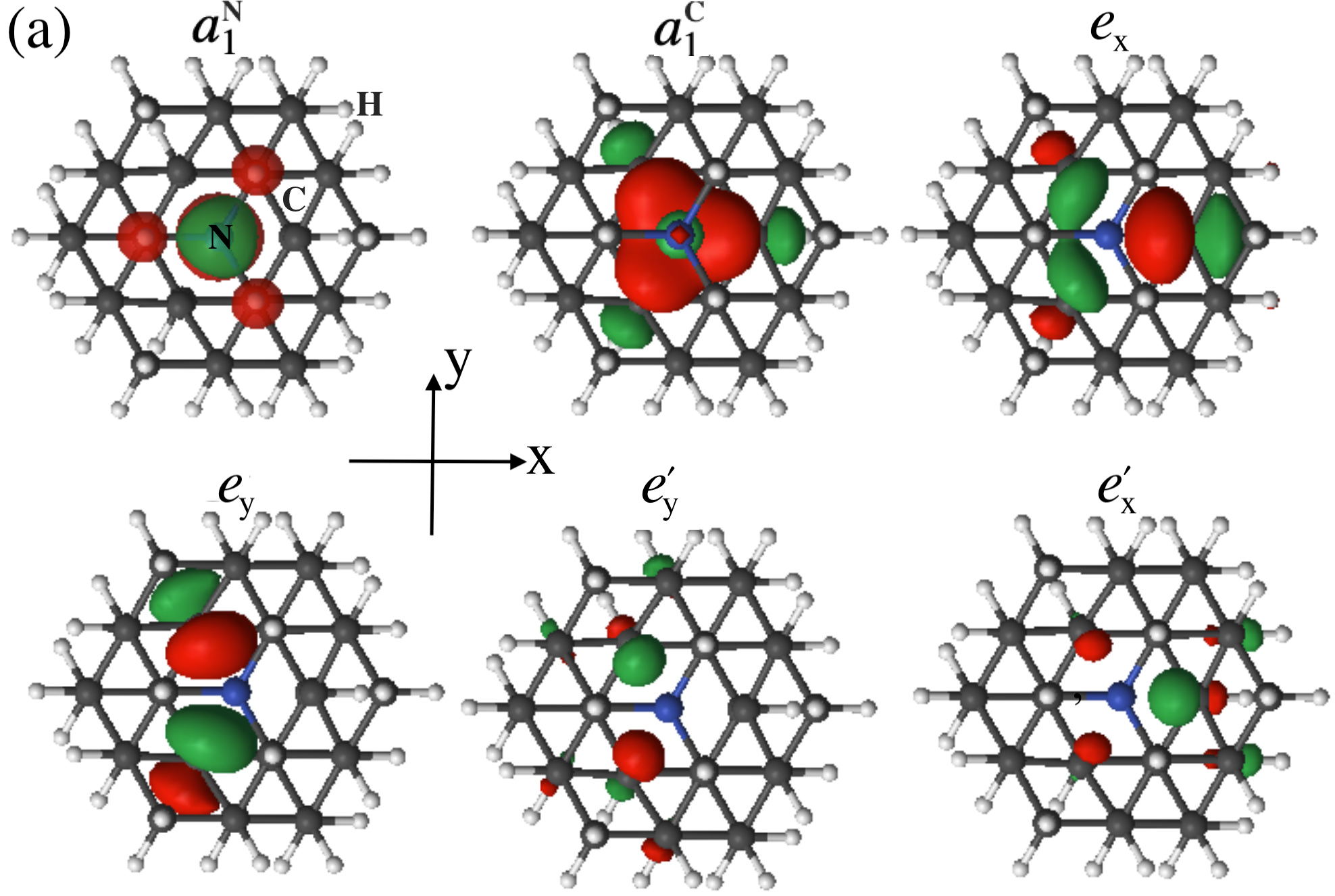}

\vspace{1.0truecm}

\includegraphics[scale=0.45]{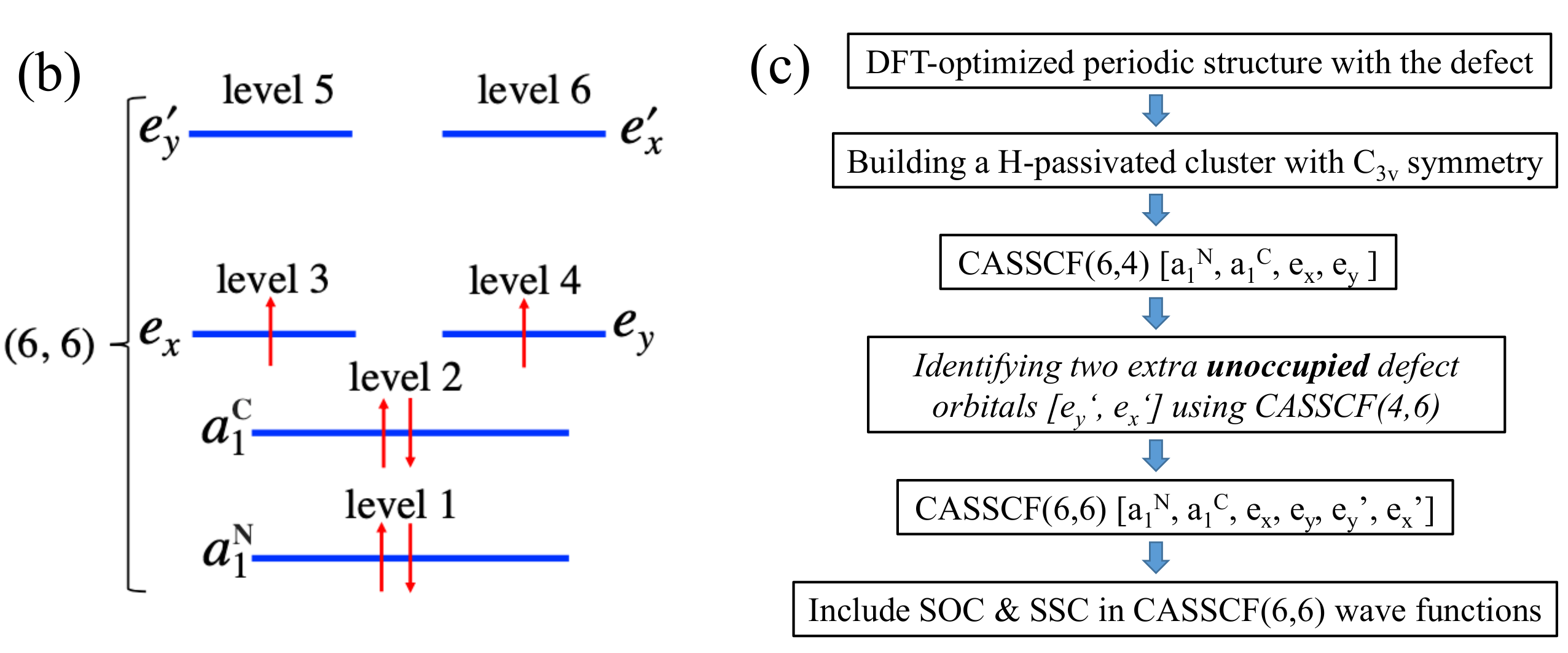}
\caption{(a) Top view of six defect orbitals (belonging to two $A_1$ and four $E$ IRRep) in the active space with iso-surface value
of 0.06 for the CASSCF(6,6) calculation of the 70-atom cluster. The similar six active orbitals are identified for the 162-atom cluster.
The LUSCUS program~\cite{luscus} is used for visualization.
(b) Nominal distribution of six electrons over the six active orbitals in the ground spin-triplet ($^3A_2$) state.
The actual occupation numbers of the $a_{1}^{N}$, $a_{1}^{C}$, $e_x$, $e_y$, $e_x^{\prime}$, $e_y^{\prime}$, are found to be
1.9986, 1.3753, 0.9883, 0.9883, 0.3248, and 0.3248, respectively, from the CASSCF(6,6) calculation.
(c) Schematic flowchart of our computational procedure in which the italicized step is discussed in detail in the Appendix.}
\label{fig2}
\end{figure*}


Inspired by the single-electron picture, we initially perform CASSCF calculations using the active space consisting of six electrons and the four dangling bond orbitals ($a_{1}^{N}$, $a_{1}^{C}$, $e_x$, and $e_y$) for the 70-atom and 162-atom diamond clusters with $C_{3v}$ symmetry,
shown in Fig.~\ref{fig2}(a) and (b). These calculations are referred to as CASSCF(6,4) following the number of electrons and orbitals used in the active space. The excitation energies obtained via CASSCF(6,4) calculations are highly overestimated as compared to experiment and the excited-state wave functions are found to be inconsistent with physical and chemical intuitions. As a result, we expand the active space by including extra unoccupied defect-localized states. The most common practice is to identify these extra states in the virtual space of the converged CASSCF(6,4) result. However, no such defect orbitals are found in the virtual space. Therefore, we introduce a series of CASSCF calculations discussed in
the Appendix (Fig.~\ref{fig:CASSCF}) in order to identify and include extra defect orbitals in the active space. With this systematic CASSCF
procedure, we find two unoccupied defect orbitals with $E$ IRRep. In order to distinguish them from the dangling bond orbitals, $e_x$ and $e_y$, they are, henceforth, referred to as $e_x^{\prime}$ and $e_y^{\prime}$ (Fig.~\ref{fig2}(a) and (b)). The $e_x^{\prime}$ and $e_y^{\prime}$ orbitals are expected to lie in the conduction band region (i.e., defect orbitals are resonant with the conduction band). With these two extra unoccupied orbitals, as well as, the four dangling bond orbitals, we form an active space consisting of six electrons and six orbitals, and carry out
CASSCF(6,6) calculations for both the total spin $S=1$ and $S=0$. Furthermore, in order to achieve high accuracy and exact numerical degeneracy
(up to $\sim 10$~neV) in states with $E$ symmetry, we carefully maintain the IRRep symmetry of all of the molecular orbitals and remove the surface-dominant orbitals in the self-consistent calculations.

\subsection{Spin-orbit coupling and spin-spin coupling}

For the ground $^3A_2$ state, the first-order SOC effect on the zero-field splitting vanishes and higher-order terms are negligibly small due to
weak SOC. However, for the first-excited $^3E$ state, the first-order SOC effect becomes important within the subspace of degenerate states
and the SOC-induced splitting turns out to be non-negligible. Therefore, for the most accurate calculation of SOC-induced splitting, we need to describe degenerate states the most accurately. In order to achieve this, state average is carried out only over the first-excited
triplet pair ($^3E$) of the CASSCF(6,6) wave functions. Then SOC is included in the converged CASSCF(6,6) spin-triplet wave functions within
the atomic mean-field approximation~\cite{Hess1996}, using the restricted active space state interaction (RASSI) method~\cite{rassi} implemented
in {\tt OpenMolcas}. For the CASSCF(6,6) energy eigenvalues and the SOC-induced zero-field splitting, {\tt OpenMolcas} is used because it provides
more accurate results due to purely symmetric orbitals and removal of surface-dominated orbitals (see the Appendix).

The zero-field splitting by the SSC is expected for all spin-triplet states. This feature is computed for the CASSCF(6,6) wave functions
using {\tt ORCA} because it is not available in {\tt OpenMolcas}. The SSC is calculated as the two-electron direct SSC over
the CASSCF(6,6) wave functions using first-order perturbation theory~\cite{Neese2006}, as implemented in {\tt ORCA}. The CASSCF(6,6) wave
functions using {\tt ORCA} are obtained by following the CASSCF procedure sketched in the Appendix without orbital symmetrization,
{\tt SUPERSYMMETRY} keyword, and removal of surface orbitals, because they are not available in {\tt ORCA}. We confirm that the zero-field
splitting induced by SSC is not sensitive to technical details of the calculations (i.e., the cluster size, the size of the active space
and the number of roots included in the state average).


\section{Results and Discussion}

\subsection{Excitation energies}

\begin{figure*}[hbt!]
\includegraphics[scale=0.75]{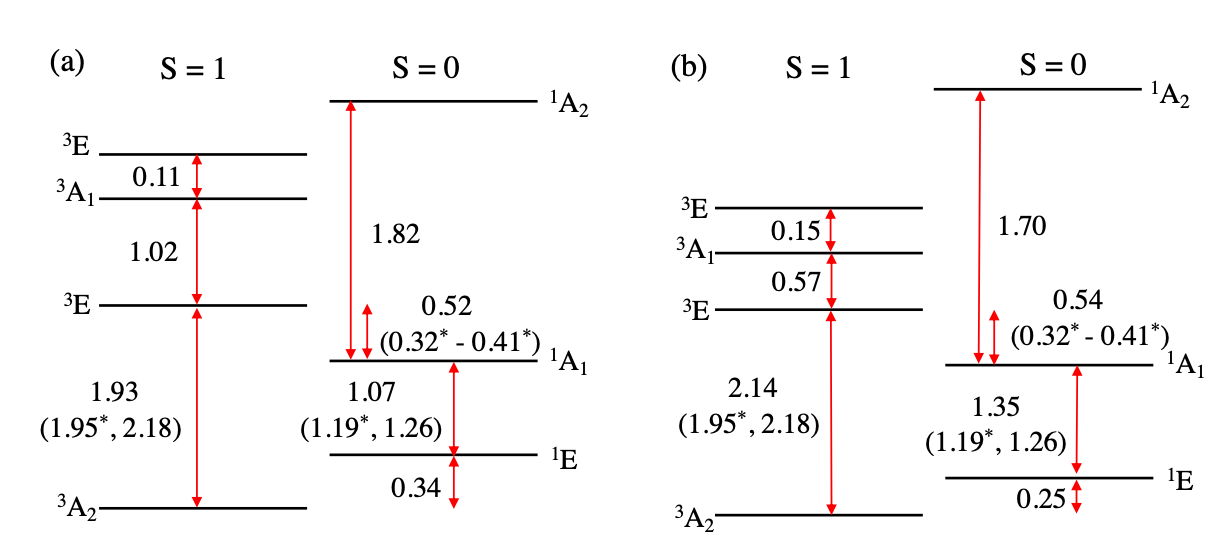}
\caption{Schematic level diagrams of the spin-triplet and spin-singlet states for (a) the 70-atom and (b) the 162-atom diamond clusters obtained
using the quantum-chemistry method without SOC or SSC. Here full electron correlation within the six molecular orbitals [Fig.~\ref{fig2}(a) and (b)]
are considered. The experimental values~\cite{RogersNJP08,Kehayias2013,Goldman2015,DaviesExpt76,Goldman2017} are shown inside parentheses.
The experimental zero-phonon absorption energies are marked with $*$. All energy values are given in units of eV.}
\label{fig3}
\end{figure*}

\begin{table*}
\begin{ruledtabular}
\caption{Our calculated excitation energies with respect to the ground state ($^3$A$_2$) in units of eV for the two cluster
sizes using the quantum chemistry method (without SOC or SSC), in comparison to previous theoretical studies and experimental data. For our calculations, neither the relaxation energy of the excited states nor vibration energies are included. In other words, we use the same geometry for the ground state and all excited triplet and singlet states. Zero-phonon absorption energies are marked with $*$. The unmarked experimental value
correspond to the vertical excitation energy, i.e. the maximum-intensity peak energy of the phonon side band spectra~\cite{DaviesExpt76}.
The experimental energy of the $^1E$ state relative to the ground-state energy is converted from the following two measurements: (a) the zero-phonon absorption energy between the $^1E$ and $^1A_1$ states which is 1.190~eV~\cite{RogersNJP08}; (b) the energy difference between the $^1A_1$ and $^3E$ states which is 0.321-0.414~eV~\cite{Goldman2015,Goldman2017}.
\label{tab1}}

\begin{tabular}{l|llllll}
Reference $\backslash$ Electronic State  &   $^3E$    &  $^3A_1$ &  $^3E$  &  $^1E$      &   $^1A_1$    &   $^1A_2$ \\
 \hline
Experiment
\cite{DaviesExpt76,Kehayias2013,RogersNJP08,Goldman2015,Goldman2017} & 1.945$^*$\cite{DaviesExpt76} &          &         & 0.34$^{*}$-0.43$^*$   & 1.51$^*$-1.60$^*$ &   \\
& $\hspace*{-0.25cm}\sim$2.18\cite{DaviesExpt76} &          &         &~\cite{Goldman2015,Goldman2017}  &~\cite{RogersNJP08} &   \\
C$_{33}$H$_{36}$N$^{\rm{-}}$ CASSCF(6,6)  &   1.93    &  2.95  &     3.06 &   0.34  &   1.41      &    3.23    \\
({\bf This work})     &           &        &          &         &             &            \\
C$_{85}$H$_{76}$N$^{\rm{-}}$ CASSCF(6,6)  &   2.14    &  2.71  &     2.86 &   0.25  &   1.60      &    3.30    \\
({\bf This work})     &           &        &          &         &             &            \\
C$_{33}$H$_{36}$N$^{\rm{-}}$ CASSCF(6,8)\cite{LinJCP08} &      2.48  &          &         &             &              &           \\
C$_{49}$H$_{52}$N$^{\rm{-}}$ CASSCF(6,8)\cite{LinJCP08} &      2.57  &          &         &             &              &           \\
C$_{19}$H$_{28}$N$^{\rm{-}}$ CASSCF(8,11)\cite{ZyubinJPC07}  &      0.98  &          &  1.22   &  0.44       &   1.00       &  1.13($^1E$) \\
C$_{19}$H$_{28}$N$^{\rm{-}}$ MRCI(8,10)\cite{ZyubinJPC07}  &      1.36  &          &  1.61   &  0.50       &   1.23       &  1.37($^1E$) \\
C$_{42}$H$_{42}$N$^{\rm{-}}$ MCCI\cite{DelaneyNL10} & 1.96, 1.93 &          &         & 0.63, 0.64  &  2.06        &          \\
$GW+$BSE~\cite{Ma2010}                              & 2.32       &          &         & 0.40        & 0.99         & 2.25($^1E'$) \\
$GW$ fit to model\cite{ChoiPRB012}   & 2.0$^*$    &     &         & $\hspace*{-0.25cm}\sim$0.5   &$\hspace*{-0.25cm}\sim$1.5    & \hspace*{-0.25cm}$\sim$3.0($^1E'$) \\
   & 2.1        &          &         &             &              &           \\
CI-CRPA\cite{Bockstedte2018}                        & 1.75$^*$ &          &         &  0.49       &   1.41       &  3.09($^1E'$) \\
(512-atom supercell)                                &  2.02    &          &         &             &              &          \\
Beyond-RPA \cite{Ma2020} with                       & 2.00     &          &         &  0.56       &   1.76       &        \\
quantum embedding theory                            &          &          &         &             &              &        \\
C$_{33}$H$_{36}$N$^{\rm{-}}$ DFT\cite{GossPRL96}    & 1.77$^*$ &     &         & 0.44        & 1.67         &           \\
DFT (512-atom                                       & 1.71$^*$ &     &         & 0.9         & 0.0, 2.2     &           \\
supercell)~\cite{GaliPRB08}                         & 1.91   &          &         &             &              &           \\
C$_{42}$H$_{42}$N$^{\rm{-}}$ DFT\cite{DelaneyNL10}       & 1.27       &          &         & 0.42        & 2.10         &          \\
                                                    &            &          &         &             & 1.26($^1A'$) &          \\
C$_{284}$H$_{144}$N$^{\rm{-}}$ DFT\cite{DelaneyNL10}     & 1.90       &          &         & 0.48        & 2.03        &          \\
                                                    &            &          &         &             & 1.26($^1A'$) &          \\
\end{tabular}
\end{ruledtabular}
\end{table*}


Figure~\ref{fig3} shows schematic level diagrams of our calculated spin-triplet and spin-singlet states for the two cluster sizes using
{\tt OpenMolcas} (quantum-chemistry methods, CASSCF(6,6)). Note that we use the ground-state geometry without phonon modes and that we do not
consider structural relaxation of the electronic excited states. An experimental absorption spectrum of an NV$^{-}$ center in diamond
consists of a sharp zero-phonon line with a broad spectrum of phonon side bands with several peaks~\cite{DaviesExpt76,Kehayias2013}. With
significant electron-phonon coupling, a zero-phonon absorption energy can noticeably differ from a vertical excitation energy. The latter energy is always higher than the former energy. The latter energy is commonly experimentally obtained from the maximum-intensity peak of the broad phonon side-band spectrum. The broadness of the phonon side bands provides some uncertainty in the maximum-intensity peak energy, which renders uncertainty
in the experimental vertical excitation energy. For comparison to experiment, we provide both experimental zero-phonon absorption energies and experimental vertical excitation energies in Fig.~\ref{fig3}.

Our calculations show that the first-excited spin-triplet $^3E$ state is separated from the ground state ($^3A_2$) by 1.93 and 2.14 eV for the
70-atom and 162-atom clusters, respectively. This energy separation does not depend much on the cluster size and it is close to the experimental
energies of zero-phonon absorption, 1.945~eV, and of vertical excitation, 2.18~eV~\cite{DaviesExpt76}.
We find that the lowest-energy singlet state has character of $^1E$ and that the first-excited singlet $^1A_1$ state is located at 1.07~eV and 1.35~eV above the $^1E$ state for the 70-atom and 162-atom clusters, respectively. The ordering and the character of the singlet states agree with experiment, considering the experimental energies of zero-phonon absorption, 1.190~eV~\cite{RogersNJP08}, and of vertical excitation, 1.26~eV~\cite{Kehayias2013}. Our results also reveal the energy differences between the triplet and singlet states. The $^1E$ state lies at 0.34~eV and 0.25~eV above the $^3A_2$ state for the 70-atom and 162-atom clusters, respectively. As a result, the energy gap between the $^3E$ and $^1A_1$ states
becomes 0.52 and 0.54~eV for the 70-atom and 162-atom clusters, respectively. Although the energy gap between the $^3A_2$ state and the $^1E$ state
has not been directly experimentally measured, the separation between the $^3E$ state and the $^1A_1$ state was measured to be 0.321-0.414~eV~\cite{Goldman2015,Goldman2017}, which is in good agreement with our results. The second-excited (third-excited) triplet state has characteristics of $^3A_1$ ($^3E$). The second-excited singlet $^1A_2$ state appears even above the third-excited triplet $^3E$ state. There are no experimental reports on the higher-energy levels or separations.


Our calculated results show that for the four lowest states ($^3A_2$, $^3E$, $^1E$, and $^1A_1$) the energy eigenvalues do not depend much on
the cluster size. However, we find that the cluster-size dependence becomes more apparent for higher-energy states, especially for the second- and third-excited triplet states ($^3A_1$ and $^3E$). Depending on the cluster size, the energy separations change but the ordering of the states does not change. A similar trend of the cluster-size dependence was reported in the complete-active space approach, using DFT Kohn-Sham orbitals and density-matrix renormalization group~\cite{Barcza2020}.
This trend can be understood by the fact that higher-energy levels have stronger electron correlations which requires inclusion of more empty
orbitals in the active space. Since experimental data are available for mainly up to the first-excited triplet $^3E$ state, we do not further
study an effect of cluster size on the electronic structure.

\subsection{Comparison to other {\it ab-initio} studies}

\begin{figure*}[hbt!]
\centering
\includegraphics[scale=0.71]{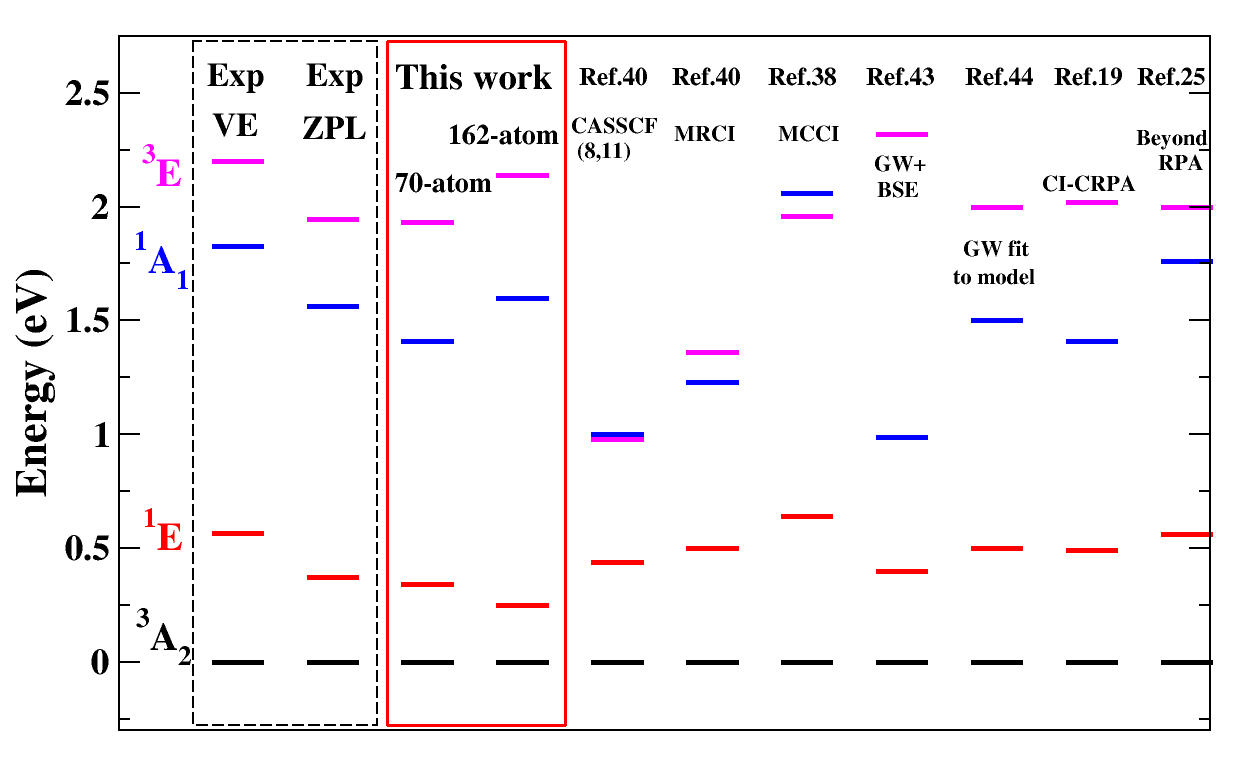}
\caption{Comparison of our calculated spin-triplet and spin-singlet energies to the previous many-body theoretical studies
\cite{Bockstedte2018,Ma2020,DelaneyNL10,ZyubinJPC07,Ma2010,ChoiPRB012}
as well as the experimental zero-phonon lines (ZPL) and vertical excitations (VE)~\cite{RogersNJP08,Kehayias2013,Goldman2015,DaviesExpt76,Goldman2017}.
The experimental VE are energies of the maximum-intensity peak of the broad phonon side-band absorption spectra.
The experimental ZPL and VE energies of the $^1E$ and $^1A_1$ states (relative to the ground state) are taken from the mid point of the
experimental range~\cite{Goldman2015} of the separation between the $^3E$ and $^1A_1$ states, while keeping the $^1A_1$$-$$^1E$ energy
difference fixed as the experimental value of 1.190~eV \cite{RogersNJP08}. }
\label{fig:summary}
\end{figure*}


Let us now compare our calculated energies of the spin-triplet and spin-singlet states ($^3A_2$, $^3E$, $^1E$, and $^1A_1$) to the
previous {\it ab-initio} theoretical studies. See Table \ref{tab1} and Fig.~\ref{fig:summary}.
In our analysis, we focus on the four lowest states because only the level separations among
them were experimentally measured and because higher-energy states are more sensitive to the cluster size and the size of the active space.
(For example, the higher-energy $^1E'$ state that many-body theory studies predicted~\cite{Bockstedte2018,Ma2010,ChoiPRB012} has
not been observed~\cite{Kehayias2013}.) We first discuss comparison to other DFT calculations and then to other quantum-chemistry studies
as well as many-body theory studies, separately.

Earlier DFT studies of an NV$^-$ center in diamond clusters and periodic supercells~\cite{GaliPRB08,Goss1996,DelaneyNL10,LinJCP08,LarsonPRB08,Galireview2019} showed that the calculated excitation energy of the $^3E$
state more or less agrees with our result and experiment except for Ref.~\cite{DelaneyNL10}. However, DFT-calculated energies of the singlet
states are scattered in a wide range and the ordering of the triplet and singlet states is inconsistent with recent experiment. This trend is
understandable considering that DFT poorly describes the singlet states due to the well-known spin contamination effect.

In the previous quantum-chemistry studies of an NV$^-$ center in diamond clusters~\cite{DelaneyNL10,ZyubinJPC07,LinJCP08}, either the excitation energies are significantly different from experiment, or the ordering of the singlet and triplet states is reversed. More specifically,
CASSCF(6,8) calculations discussed in Ref.~\cite{LinJCP08} showed that the excitation energy of the ${^3}$E state is 0.5-0.6~eV (0.3-0.4 eV)
higher than our result (experiment). The singlet states were not investigated in that work. In the CASSCF(8,11) calculations presented in Ref.~\cite{ZyubinJPC07},  the excitation energy of the ${^3}$E state is about 1.0 eV lower than our result or experiment, and the singlet $^1A_1$ state is slightly above the triplet $^3E$ state, which does not agree with our result or recent experiment. Their multireference configuration interaction (MRCI) calculations~\cite{ZyubinJPC07} somewhat increase the energies of the triplet and singlet states with the correct ordering of the excited triplet and singlet states. However, the energy of the $^3E$ state remains lower than our value by about 0.6~eV. Monte Carlo configuration interaction (MCCI) studies~\cite{DelaneyNL10} showed the energy of the $^3E$ state in agreement with our result and experiment. However, the ordering of the $^3E$ and $^1A_1$ states is reversed. See Table~\ref{tab1} and Fig.~\ref{fig:summary}. The discrepancies between our results and all of the earlier quantum-chemistry calculations arise from the choice of orbitals in the active space. One of the most common ways to choose active orbitals is to
use single-electron molecular orbitals in the vicinity of the band gap such as orbitals near the highest occupied molecular orbitals (HOMO) and lowest unoccupied molecular orbitals (LUMO). For a hydrogen-passivated diamond cluster with an NV$^{-}$ center, either this common practice within CASSCF or MRCI, or an automatic choice of the active space in MCCI may result in non-physical surface-dominated orbitals in the CI basis set.
As shown in Fig.~\ref{fig2}(a), in our case, all six orbitals in the active are localized near the vacancy defect.


An earlier many-body perturbation study~\cite{Ma2010} based on the $GW$ approximation with Bethe-Salpeter equation (BSE) provided the singlet-singlet ($^1A_1$ - $^1E$) energy difference about 0.6~eV lower than our result and recent experiment~\cite{Kehayias2013,Goldman2015}, although the energy of the $^1E$ state relative to the ground state, as well as the ordering of the two singlet states are in agreement with the recent experimental data.
On the other hand, recent many-body studies~\cite{Bockstedte2018,Ma2020,ChoiPRB012} showed more promising results by using effective many-body model Hamiltonians with parameters obtained from (or fitted to) {\it ab-initio} calculations in order to properly include many-body character in the wave functions. For example, additional unoccupied defect states (resonant to the conduction band) and doubly occupied defect states (in the valence band) were included in the configuration interaction constrained random phase approximation (CI-CRPA) method~\cite{Bockstedte2018}. This is analogous to our inclusion of unoccupied level 5 and 6 and doubly occupied level 1 [Fig.~\ref{fig2}(b)] in the active space for proper treatment of electron correlation. Their results are closest to our result among the previous studies that we have discussed (see Fig.~\ref{fig:summary} and Table~\ref{tab1}). Yet, there are some differences. In the fitting of $GW$-calculated bands to model Hamiltonian~\cite{ChoiPRB012} (in the
CI-CRPA method~\cite{Bockstedte2018}), the singlet-singlet energy difference is about 0.2-0.3 eV (0.3-0.4~eV) lower than our result and experiment.
In the beyond-RPA implemented in the quantum embedding theory \cite{Ma2020}, the energy difference between the $^3E$ and $^1A_1$ states is somewhat smaller than our result and experiment. This discrepancy may arise from missing orbital configurations in the $^1E$ and $^1A_1$ states in Refs.~\cite{Bockstedte2018,Ma2020,ChoiPRB012} that are discussed in Sec.\ref{eigenstates}. Here we stress that it does not seem to be straightforward to include effects of SOC and SSC within the formalisms used in Refs.~\cite{Bockstedte2018,Ma2020,ChoiPRB012} in contrast to the quantum chemistry methods where such effects can be added to the many-body wave functions without an introduction of new fitting parameters (see Sec.\ref{ZeroField}).

\subsection{Characteristics of energy eigenstates}\label{eigenstates}

\begin{table*}
\begin{ruledtabular}
\caption{Characteristics of the calculated energy eigenstates for the 70-atom cluster using the configuration (spin-free) basis states. Here the
configuration basis states are all possible states generating the maximum $M_z$ value from the six active orbitals (Fig.~\ref{fig2}) for a given
total spin $S$, where $M_z$ is an eigenvalue of $S_z$. Each box represents an orbital. Up and down arrows denote spin-up and spin-down electrons.
Each configuration represents a Slater's determinant of the orbitals with $2S+1$ degeneracy. Percentages denote orbital configuration weights.
Only configurations with weights greater than 5\% or above are listed. Weights greater than 10\% are denoted as boldface.
\label{tab:conf}}
\begin{tabular}{l | l}
State & Configuration (weight) \framebox[0.2in] {$~a_{1{\color{white}y}}^{N{\color{white}'}}$}\framebox[0.2in] {$~a_{1{\color{white}y}}^{C{\color{white}'}}$}\framebox[0.2in] {$~e_{x{\color{white}y}}^{{\color{white}'}}$}\framebox[0.2in] {$~e_y^{{\color{white}C'}}$}\framebox[0.2in] {$~e_{y{\color{white}y}}^{'}$}\framebox[0.2in] {$~e_{x{\color{white}y}}^{'}$} \\ \hline
$^3A_2 (\Psi_{1,T})$ & {\tiny\framebox[0.125in] {${\color{black} \uparrow\downarrow}$}\hspace*{-0.098cm} \framebox[0.125in] {${\color{black}\uparrow \downarrow}$}\hspace*{-0.098cm} \framebox[0.125in] {$~~{\color{black}\uparrow} {\color{white}\uparrow}$}\hspace*{-0.098cm} \framebox[0.125in] {$~~{\color{black}\uparrow} {\color{white}\uparrow}$}\hspace*{-0.098cm} \framebox[0.125in] {${\color{white}\uparrow} {\color{white}\uparrow}$}\hspace*{-0.098cm} \framebox[0.125in] {${\color{white}\uparrow} {\color{white}\uparrow}$}} (\bf 94\%) \\\\

$^3E ~(\Psi_{2,T})$ & {\tiny  \framebox[0.125in] {${\color{black}\uparrow \downarrow}$}\hspace*{-0.098cm} \framebox[0.125in] {$~~{\color{black}\uparrow}{\color{white}\uparrow}$}\hspace*{-0.098cm} \framebox[0.125in] {${\color{black}\uparrow \downarrow}$} \hspace*{-0.186cm} \framebox[0.125in] {$~~{\color{black}\uparrow} {\color{white}\uparrow}$}\hspace*{-0.098cm} \framebox[0.125in] {${\color{white}\uparrow} {\color{white}\uparrow}$}\hspace*{-0.098cm} \framebox[0.125in] {${\color{white}\uparrow} {\color{white}\uparrow}$}} ({\bf 38\%}),
{\tiny  \framebox[0.125in] {${\color{black}\uparrow \downarrow}$}\hspace*{-0.098cm} \framebox[0.125in] {$~~{\color{black}\uparrow} {\color{white}\uparrow}$}\hspace*{-0.098cm} \framebox[0.125in] {$~~{\color{black}\uparrow}  {\color{white}\uparrow}$} \hspace*{-0.189cm} \framebox[0.125in] {${\color{black}\uparrow \downarrow}$}\hspace*{-0.098cm} \framebox[0.125in] {${\color{white}\uparrow} {\color{white}\uparrow}$}\hspace*{-0.098cm} \framebox[0.125in] {${\color{white}\uparrow} {\color{white}\uparrow}$}} ({\bf 31\%}),
{\tiny  \framebox[0.125in] {${\color{black}\uparrow \downarrow}$}\hspace*{-0.098cm} \framebox[0.125in] {$~~{\color{black}\uparrow} {\color{white}\uparrow}$}\hspace*{-0.098cm} \framebox[0.125in] {$~~{\color{black}\uparrow}  {\color{white}\uparrow}$} \hspace*{-0.186cm} \framebox[0.125in] {$~~{\color{black}\uparrow}{\color{white}\uparrow}$}\hspace*{-0.098cm} \framebox[0.125in] {${\color{white}\uparrow} {\color{white}\uparrow}$}\hspace*{-0.098cm} \framebox[0.125in] {$~~{\color{black}\downarrow}{\color{white}\downarrow}$}} (7\%),
{\tiny\hspace*{0.09cm}  \framebox[0.125in] {${\color{black}\uparrow \downarrow}$}\hspace*{-0.098cm} \framebox[0.125in] {$~~{\color{black}\uparrow} {\color{white}\uparrow}$}\hspace*{-0.098cm} \framebox[0.125in] {$~~{\color{black}\uparrow}  {\color{white}\uparrow}$} \hspace*{-0.189cm} \framebox[0.125in] {$~~{\color{black}\uparrow}{\color{white}\uparrow}$}\hspace*{-0.098cm} \framebox[0.125in] {$~~~{\color{black}\downarrow}{\color{white}\downarrow} {\color{white}\uparrow}$}\hspace*{-0.098cm} \framebox[0.125in] {${\color{white}\uparrow}{\color{white}\uparrow}$}} (5\%) \\\\

        ~~~~~($\Psi_{3,T})$ & {\tiny  \framebox[0.125in] {${\color{black}\uparrow \downarrow}$}\hspace*{-0.098cm} \framebox[0.125in] {$~~{\color{black}\uparrow}{\color{white}\uparrow} $}\hspace*{-0.098cm} \framebox[0.125in] {$~~{\color{black}\uparrow} {\color{white}\uparrow}$}\hspace*{-0.098cm} \framebox[0.125in] {${\color{black}\uparrow \downarrow}$}\hspace*{-0.098cm} \framebox[0.125in] {${\color{white}\uparrow} {\color{white}\uparrow}$}\hspace*{-0.098cm} \framebox[0.125in] {${\color{white}\uparrow} {\color{white}\uparrow}$}} ({\bf 38\%}),
{\tiny  \framebox[0.125in] {${\color{black}\uparrow \downarrow}$}\hspace*{-0.098cm} \framebox[0.125in] {$~~{\color{black}\uparrow}{\color{white}\uparrow}$}\hspace*{-0.098cm} \framebox[0.125in] {${\color{black}\uparrow \downarrow}$} \hspace*{-0.186cm} \framebox[0.125in] {$~~{\color{black}\uparrow} {\color{white}\uparrow}$}\hspace*{-0.098cm} \framebox[0.125in] {${\color{white}\uparrow} {\color{white}\uparrow}$}\hspace*{-0.098cm} \framebox[0.125in] {${\color{white}\uparrow} {\color{white}\uparrow}$}} ({\bf 30\%}),
{\tiny  \framebox[0.125in] {${\color{black}\uparrow \downarrow}$}\hspace*{-0.098cm} \framebox[0.125in] {$~~{\color{black}\uparrow} {\color{white}\uparrow}$}\hspace*{-0.098cm} \framebox[0.125in] {$~~{\color{black}\uparrow}  {\color{white}\uparrow}$} \hspace*{-0.186cm} \framebox[0.125in] {$~~{\color{black}\uparrow}{\color{white}\uparrow}$}\hspace*{-0.098cm} \framebox[0.125in] {$~~~{\color{black}\downarrow}{\color{white}\downarrow} {\color{white}\uparrow}$}\hspace*{-0.098cm} \framebox[0.125in] {${\color{white}\uparrow}{\color{white}\uparrow}$}} (7\%),
{\tiny\hspace*{0.09cm}  \framebox[0.125in] {${\color{black}\uparrow \downarrow}$}\hspace*{-0.098cm} \framebox[0.125in] {$~~{\color{black}\uparrow} {\color{white}\uparrow}$}\hspace*{-0.098cm} \framebox[0.125in] {$~~\uparrow  {\color{white}\uparrow}$} \hspace*{-0.186cm} \framebox[0.125in] {$~~\uparrow{\color{white}\uparrow}$}\hspace*{-0.098cm} \framebox[0.125in] {${\color{white}\uparrow} {\color{white}\uparrow}$}\hspace*{-0.098cm} \framebox[0.125in] {$~~{\color{black}\downarrow}{\color{white}\downarrow}$}} (5\%) \\\\

 $^3A_1 (\Psi_{4,T})$ &{\tiny  \framebox[0.125in] {${\color{black}\uparrow \downarrow}$}\hspace*{-0.098cm} \framebox[0.125in] {${\color{black}\uparrow \downarrow}$}\hspace*{-0.098cm} \framebox[0.125in] {$~~{\color{black}\uparrow} {\color{white}\uparrow}$}\hspace*{-0.098cm} \framebox[0.125in] {${\color{white}\uparrow} {\color{white}\uparrow}$}\hspace*{-0.098cm} \framebox[0.125in] {$ {\color{white}\uparrow} {\color{white}\uparrow}$}\hspace*{-0.098cm} \framebox[0.125in] {$~~{\color{black}\uparrow} {\color{white}\uparrow}$}} ({\bf 29\%}),

{\tiny  \framebox[0.125in] {$\uparrow \downarrow$}\hspace*{-0.098cm} \framebox[0.125in] {$\uparrow \downarrow$}\hspace*{-0.098cm} \framebox[0.125in] {${\color{white}\uparrow} {\color{white}\uparrow}$}\hspace*{-0.098cm} \framebox[0.125in] {$~~\uparrow {\color{white}\uparrow}$}\hspace*{-0.098cm} \framebox[0.125in] {$~~\uparrow {\color{white}\uparrow}$}\hspace*{-0.098cm} \framebox[0.125in] {${\color{white}\uparrow} {\color{white}\uparrow}$}} ({\bf 29\%}),
{\tiny  \framebox[0.125in] {$\uparrow \downarrow$}\hspace*{-0.098cm} \framebox[0.125in] {$~~\uparrow {\color{white}\uparrow}$}\hspace*{-0.098cm} \framebox[0.125in] {$~~\uparrow {\color{white}\uparrow}$}\hspace*{-0.098cm} \framebox[0.125in] {$~~\downarrow {\color{white}\uparrow}$}\hspace*{-0.098cm} \framebox[0.125in] {${\color{white}\uparrow} {\color{white}\uparrow}$}\hspace*{-0.098cm} \framebox[0.125in] {$~~\uparrow {\color{white}\uparrow}$}} (9\%),
         {\tiny  \framebox[0.125in] {$\uparrow \downarrow$}\hspace*{-0.098cm} \framebox[0.125in] {$~~\uparrow {\color{white}\uparrow}$}\hspace*{-0.098cm} \framebox[0.125in] {$\uparrow \downarrow$}\hspace*{-0.098cm} \framebox[0.125in] {${\color{white}\uparrow} {\color{white}\uparrow}$}\hspace*{-0.098cm} \framebox[0.125in] {$~~\uparrow {\color{white}\uparrow}$}\hspace*{-0.098cm} \framebox[0.125in] {${\color{white}\uparrow} {\color{white}\uparrow}$}} (6\%),

{\tiny \hspace*{0.176cm}\framebox[0.125in] {$\uparrow \downarrow$}\hspace*{-0.098cm} \framebox[0.125in] {$~~\uparrow {\color{white}\uparrow}$}\hspace*{-0.098cm} \framebox[0.125in] {${\color{white}\uparrow} {\color{white}\uparrow}$}\hspace*{-0.098cm} \framebox[0.125in]{$\uparrow \downarrow$}\hspace*{-0.098cm} \framebox[0.125in] {$~~\uparrow {\color{white}\uparrow}$}\hspace*{-0.098cm} \framebox[0.125in] {${\color{white}\uparrow} {\color{white}\uparrow}$}} (6\%),\\\\ &
{\tiny \framebox[0.125in] {$\uparrow \downarrow$}\hspace*{-0.098cm} \framebox[0.125in]{${\color{white}\uparrow} {\color{white}\uparrow}$}\hspace*{-0.098cm} \framebox[0.125in]{$\uparrow \downarrow$}\hspace*{-0.098cm} \framebox[0.125in] {$~~\uparrow {\color{white}\uparrow}$}\hspace*{-0.098cm} \framebox[0.125in] {$~~\uparrow {\color{white}\uparrow}$}\hspace*{-0.098cm} \framebox[0.125in]{${\color{white}\uparrow} {\color{white}\uparrow}$}} (6\%),
{\tiny\hspace*{0.176cm}\framebox[0.125in] {$\uparrow \downarrow$}\hspace*{-0.098cm} \framebox[0.125in]{${\color{white}\uparrow} {\color{white}\uparrow}$}\hspace*{-0.098cm} \framebox[0.125in]{$~~\uparrow {\color{white}\uparrow}$}\hspace*{-0.098cm} \framebox[0.125in]{$\uparrow \downarrow$}\hspace*{-0.098cm} \framebox[0.125in]{${\color{white}\uparrow} {\color{white}\uparrow}$}\hspace*{-0.098cm} \framebox[0.125in]{$~~\uparrow {\color{white}\uparrow}$}} (6\%) \\\\

$^3E ~(\Psi_{5,T})$ &{\tiny  \framebox[0.125in] {$\uparrow \downarrow$}\hspace*{-0.098cm} \framebox[0.125in] {$\uparrow \downarrow$}\hspace*{-0.098cm} \framebox[0.125in] {$~~\uparrow {\color{white}\uparrow}$}\hspace*{-0.098cm} \framebox[0.125in] {${\color{white}\uparrow} {\color{white}\uparrow}$}\hspace*{-0.098cm} \framebox[0.125in] {$~~\uparrow {\color{white}\uparrow}$}\hspace*{-0.098cm} \framebox[0.125in] {${\color{white}\uparrow} {\color{white}\uparrow}$}} ({\bf22\%}),

{\tiny  \framebox[0.125in] {$\uparrow \downarrow$}\hspace*{-0.098cm} \framebox[0.125in] {$\uparrow \downarrow$}\hspace*{-0.098cm} \framebox[0.125in] {${\color{white}\uparrow}{\color{white}\uparrow}$}\hspace*{-0.098cm} \framebox[0.125in] {$~~\uparrow {\color{white}\uparrow}$}\hspace*{-0.098cm} \framebox[0.125in] {${\color{white}\uparrow}{\color{white}\uparrow}$}\hspace*{-0.098cm} \framebox[0.125in] {$~~\uparrow {\color{white}\uparrow}$}} ({\bf22\%}),

{\tiny  \framebox[0.125in] {$\uparrow \downarrow$}\hspace*{-0.098cm} \framebox[0.125in] {$~~\uparrow {\color{white}\uparrow}$}\hspace*{-0.098cm} \framebox[0.125in] {$~~\uparrow{\color{white}\uparrow}$}\hspace*{-0.098cm} \framebox[0.125in] {$~~\uparrow {\color{white}\uparrow}$}\hspace*{-0.098cm} \framebox[0.125in] {$~~\downarrow{\color{white}\uparrow}$}\hspace*{-0.098cm} \framebox[0.125in] {${\color{white}\uparrow} {\color{white}\uparrow}$}} ({\bf14\%}),

{\tiny  \framebox[0.125in] {$\uparrow \downarrow$}\hspace*{-0.098cm} \framebox[0.125in] {$~~\uparrow {\color{white}\uparrow}$}\hspace*{-0.098cm} \framebox[0.125in] {$~~\uparrow{\color{white}\uparrow}$}\hspace*{-0.098cm} \framebox[0.125in] {$~~\downarrow {\color{white}\uparrow}$}\hspace*{-0.098cm} \framebox[0.125in] {$~~\uparrow{\color{white}\uparrow}$}\hspace*{-0.098cm} \framebox[0.125in] {${\color{white}\uparrow} {\color{white}\uparrow}$}} (6\%),

{\tiny  \framebox[0.125in] {$\uparrow \downarrow$}\hspace*{-0.098cm} \framebox[0.125in] {$~~\uparrow {\color{white}\uparrow}$}\hspace*{-0.098cm} \framebox[0.125in] {$~~\uparrow{\color{white}\uparrow}$}\hspace*{-0.098cm} \framebox[0.125in] {$~~\uparrow {\color{white}\uparrow}$}\hspace*{-0.098cm} \framebox[0.125in] {${\color{white}\uparrow}{\color{white}\uparrow}$}\hspace*{-0.098cm} \framebox[0.125in] {$~~\downarrow {\color{white}\uparrow}$}} (5\%) \\\\

 ~~~~~($\Psi_{6,T}$) &{\tiny  \framebox[0.125in] {$\uparrow \downarrow$}\hspace*{-0.098cm} \framebox[0.125in] {$\uparrow \downarrow$}\hspace*{-0.098cm} \framebox[0.125in] {$~~\uparrow{\color{white}\uparrow}$}\hspace*{-0.098cm} \framebox[0.125in] {${\color{white}\uparrow}{\color{white}\uparrow}$}\hspace*{-0.098cm} \framebox[0.125in] {${\color{white}\uparrow} {\color{white}\uparrow}$}\hspace*{-0.098cm} \framebox[0.125in] {$~~\uparrow{\color{white}\uparrow}$}} ({\bf 22\%}),

{\tiny  \framebox[0.125in] {$\uparrow \downarrow$}\hspace*{-0.098cm} \framebox[0.125in] {$\uparrow \downarrow$}\hspace*{-0.098cm} \framebox[0.125in] {${\color{white}\uparrow}{\color{white}\uparrow}$}\hspace*{-0.098cm} \framebox[0.125in] {$~~\uparrow {\color{white}\uparrow}$}\hspace*{-0.098cm} \framebox[0.125in] {$~~\uparrow{\color{white}\uparrow}$}\hspace*{-0.098cm} \framebox[0.125in] {${\color{white}\uparrow} {\color{white}\uparrow}$}} ({\bf 22\%}),

{\tiny  \framebox[0.125in] {$\uparrow \downarrow$}\hspace*{-0.098cm} \framebox[0.125in] {$~~\uparrow {\color{white}\uparrow}$}\hspace*{-0.098cm} \framebox[0.125in] {$~~\uparrow{\color{white}\uparrow}$}\hspace*{-0.098cm} \framebox[0.125in] {$~~\uparrow {\color{white}\uparrow}$}\hspace*{-0.098cm} \framebox[0.125in] {${\color{white}\uparrow}{\color{white}\uparrow}$}\hspace*{-0.098cm} \framebox[0.125in] {$~~\downarrow{\color{white}\uparrow}$}} ({\bf14\%}),

{\tiny  \framebox[0.125in] {$\uparrow \downarrow$}\hspace*{-0.098cm} \framebox[0.125in] {$~~\uparrow {\color{white}\uparrow}$}\hspace*{-0.098cm} \framebox[0.125in] {$~~\uparrow{\color{white}\uparrow}$}\hspace*{-0.098cm} \framebox[0.125in] {$~~\downarrow {\color{white}\uparrow}$}\hspace*{-0.098cm} \framebox[0.125in] {${\color{white}\uparrow}{\color{white}\uparrow}$}\hspace*{-0.098cm} \framebox[0.125in] {$~~\uparrow {\color{white}\uparrow}$}} (6\%)

{\tiny  \framebox[0.125in] {$\uparrow \downarrow$}\hspace*{-0.098cm} \framebox[0.125in] {$~~\uparrow {\color{white}\uparrow}$}\hspace*{-0.098cm} \framebox[0.125in] {$~~\uparrow{\color{white}\uparrow}$}\hspace*{-0.098cm} \framebox[0.125in] {$~~\uparrow {\color{white}\uparrow}$}\hspace*{-0.098cm} \framebox[0.125in] {$~~\downarrow{\color{white}\uparrow}$}\hspace*{-0.098cm} \framebox[0.125in] {${\color{white}\uparrow} {\color{white}\uparrow}$}} (5\%) \\\\
 \hline
$^1E~ (\Psi_{1,S})$ &{\tiny  \framebox[0.125in] {$\uparrow \downarrow$}\hspace*{-0.098cm} \framebox[0.125in]{$\uparrow \downarrow$} \hspace*{-0.186cm} \framebox[0.125in] {$\uparrow \downarrow$}\hspace*{-0.098cm} \framebox[0.125in] {${\color{white}\uparrow} {\color{white}\uparrow}$}\hspace*{-0.098cm} \framebox[0.125in] {${\color{white}\uparrow} {\color{white}\uparrow}$}\hspace*{-0.098cm} \framebox[0.125in] {${\color{white}\uparrow} {\color{white}\uparrow}$}} ({\bf 34\%}),
 {\tiny  \framebox[0.125in] {$\uparrow \downarrow$}\hspace*{-0.098cm} \framebox[0.125in] {$\uparrow \downarrow$} \hspace*{-0.186cm} \framebox[0.125in] {${\color{white}\uparrow}{\color{white}\uparrow}$}\hspace*{-0.098cm} \framebox[0.125in] {$\uparrow \downarrow$}\hspace*{-0.098cm} \framebox[0.125in] {${\color{white}\uparrow} {\color{white}\uparrow}$}\hspace*{-0.098cm} \framebox[0.125in] {${\color{white}\uparrow} {\color{white}\uparrow}$}} ({\bf34\%}),
 {\tiny  \framebox[0.125in] {$\uparrow \downarrow$}\hspace*{-0.098cm} \framebox[0.125in] {$~~\uparrow {\color{white}\uparrow}$} \hspace*{-0.186cm} \framebox[0.125in] {$~~\downarrow {\color{white}\uparrow}$}\hspace*{-0.098cm} \framebox[0.125in] {$\uparrow \downarrow$}\hspace*{-0.098cm} \framebox[0.125in] {${\color{white}\uparrow} {\color{white}\uparrow}$}\hspace*{-0.098cm} \framebox[0.125in] {${\color{white}\uparrow} {\color{white}\uparrow}$}} ({\bf 12\%}),
{\tiny \framebox[0.125in] {$\uparrow \downarrow$}\hspace*{-0.098cm} \framebox[0.125in] {$\uparrow \downarrow$}\hspace*{-0.098cm} \framebox[0.125in] {$~~\uparrow {\color{white}\uparrow}$} \hspace*{-0.186cm} \framebox[0.125in] {$~~\downarrow {\color{white}\uparrow}$}\hspace*{-0.098cm} \framebox[0.125in] {${\color{white}\uparrow} {\color{white}\uparrow}$}\hspace*{-0.098cm} \framebox[0.125in] {${\color{white}\uparrow} {\color{white}\uparrow}$}} (7\%) \\\\
 ~~~~~($\Psi_{2,S}$) &{\tiny  \framebox[0.125in] {$\uparrow \downarrow$}\hspace*{-0.098cm} \framebox[0.125in] {$\uparrow \downarrow$}\hspace*{-0.1cm} \framebox[0.125in] {$~~\uparrow {\color{white}\uparrow}$} \hspace*{-0.18cm} \framebox[0.125in] {$~~\downarrow {\color{white}\uparrow}$}\hspace*{-0.098cm} \framebox[0.125in] {${\color{white}\uparrow} {\color{white}\uparrow}$}\hspace*{-0.098cm} \framebox[0.125in] {${\color{white}\uparrow} {\color{white}\uparrow}$}} ({\bf69\%}),
{\tiny  \framebox[0.125in] {$\uparrow \downarrow$}\hspace*{-0.098cm} \framebox[0.125in] {$~~\uparrow{\color{white}\uparrow}$}\hspace*{-0.098cm} \framebox[0.125in] {$\uparrow \downarrow$} \hspace*{-0.186cm} \framebox[0.125in] {$~~\downarrow {\color{white}\uparrow}$}\hspace*{-0.098cm} \framebox[0.125in] {${\color{white}\uparrow} {\color{white}\uparrow}$}\hspace*{-0.098cm} \framebox[0.125in] {${\color{white}\uparrow} {\color{white}\uparrow}$}} ({\bf 12\%}) \\\\

$^1A_1 (\Psi_{3,S})$ & {\tiny  \framebox[0.125in] {$\uparrow \downarrow$}\hspace*{-0.098cm} \framebox[0.125in]{$\uparrow \downarrow$} \hspace*{-0.18cm} \framebox[0.125in] {$\uparrow \downarrow$}\hspace*{-0.098cm} \framebox[0.125in] {${\color{white}\uparrow} {\color{white}\uparrow}$}\hspace*{-0.098cm} \framebox[0.125in] {${\color{white}\uparrow} {\color{white}\uparrow}$}\hspace*{-0.098cm} \framebox[0.125in] {${\color{white}\uparrow} {\color{white}\uparrow}$}} ({\bf29\%}),
 {\tiny  \framebox[0.125in] {$\uparrow \downarrow$}\hspace*{-0.098cm} \framebox[0.125in] {$\uparrow \downarrow$} \hspace*{-0.186cm} \framebox[0.125in] {${\color{white}\uparrow}{\color{white}\uparrow}$}\hspace*{-0.1cm} \framebox[0.125in] {$\uparrow \downarrow$}\hspace*{-0.098cm} \framebox[0.125in] {${\color{white}\uparrow} {\color{white}\uparrow}$}\hspace*{-0.098cm} \framebox[0.125in] {${\color{white}\uparrow} {\color{white}\uparrow}$}} ({\bf29\%}),
 {\tiny  \framebox[0.125in] {$\uparrow \downarrow$}\hspace*{-0.098cm} \framebox[0.125in] {${\color{white}\uparrow}{\color{white}\uparrow}$} \hspace*{-0.186cm} \framebox[0.125in] {$\uparrow \downarrow$}\hspace*{-0.098cm} \framebox[0.125in] {$\uparrow \downarrow$}\hspace*{-0.098cm} \framebox[0.125in] {${\color{white}\uparrow} {\color{white}\uparrow}$}\hspace*{-0.098cm} \framebox[0.125in] {${\color{white}\uparrow} {\color{white}\uparrow}$}} ({\bf20\%}) \\\\

$^1A_2 (\Psi_{4,S})$ &  {\tiny  \framebox[0.125in] {$\uparrow \downarrow$}\hspace*{-0.098cm} \framebox[0.125in]{$\uparrow \downarrow$} \hspace*{-0.186cm} \framebox[0.125in] {$~\uparrow {\color{white}\uparrow}$}\hspace*{-0.098cm} \framebox[0.125in] {$ {\color{white}\uparrow} {\color{white}\uparrow}$}\hspace*{-0.098cm} \framebox[0.125in] {$~~\downarrow{\color{white}\uparrow}$}\hspace*{-0.098cm} \framebox[0.125in] {${\color{white}\uparrow}{\color{white}\uparrow}$}} ({\bf32\%}),

 {\tiny  \framebox[0.125in] {$\uparrow \downarrow$}\hspace*{-0.098cm} \framebox[0.125in] {$\uparrow \downarrow$}\hspace*{-0.098cm} \framebox[0.125in] {${\color{white}\uparrow} {\color{white}\uparrow}$} \hspace*{-0.186cm} \framebox[0.125in] {$~~\uparrow{\color{white}\uparrow}$}\hspace*{-0.098cm} \framebox[0.125in] {${\color{white}\uparrow} {\color{white}\uparrow}$}\hspace*{-0.098cm} \framebox[0.125in] {$~~\downarrow {\color{white}\uparrow}$}} ({\bf32\%}),

 {\tiny  \framebox[0.125in] {$\uparrow \downarrow$}\hspace*{-0.098cm} \framebox[0.125in] {${\color{white}\uparrow} {\color{white}\uparrow}$}\hspace*{-0.098cm} \framebox[0.125in] {$~~\uparrow {\color{white}\uparrow} $} \hspace*{-0.186cm} \framebox[0.125in] {$\uparrow \downarrow$}\hspace*{-0.098cm} \framebox[0.125in] {$~~\downarrow {\color{white}\uparrow}$}\hspace*{-0.098cm} \framebox[0.125in] {${\color{white}\uparrow} {\color{white}\uparrow}$}} (6\%),

{\tiny\hspace*{0.09cm}  \framebox[0.125in] {$\uparrow \downarrow$}\hspace*{-0.098cm} \framebox[0.125in] {${\color{white}\uparrow} {\color{white}\uparrow}$}\hspace*{-0.098cm} \framebox[0.125in] {$\uparrow \downarrow$}\hspace*{-0.098cm} \framebox[0.125in] {$~~\uparrow {\color{white}\uparrow}$} \hspace*{-0.186cm} \framebox[0.125in] {${\color{white}\uparrow} {\color{white}\uparrow}$}\hspace*{-0.098cm} \framebox[0.125in] {$~~\downarrow{\color{white}\uparrow}$}} (6\%) \\
\end{tabular}
\end{ruledtabular}
\end{table*}


We now discuss characteristics of our calculated triplet and singlet energy eigenstates (Table~\ref{tab:conf}).
Here we use configuration basis states which are all possible states generating the maximum $M_z$ value from the six active orbitals for a given
total spin $S$, where $M_z$ is an eigenvalue of the $S_z$ operator. The total wave functions in terms of true $S_z$ eigenstates are obtained when
SOC is applied to the many-body (CASSCF) wave functions within the RASSI method~\cite{rassi} using the Wigner-Eckart theorem. The SOC effect is discussed later in Sec.~\ref{ZeroField}.

For the ground and first-excited triplet states, the configurations of our calculated eigenstates are similar to those identified from the phenomenological molecular models based on group theory~\cite{MazeNJP011,DohertyNJP011,Tamarat2008}, as long as we focus on the configurations with weights greater than 10\%. However, for the {\it singlet} states, we find that the following additional configurations significantly contribute:
$ {\tiny  \framebox[0.125in] {$\uparrow \downarrow$}\hspace*{-0.0cm} \framebox[0.125in] {$~~\uparrow {\color{white}\uparrow}$} \hspace*{-0cm} \framebox[0.125in] {$~~\downarrow {\color{white}\uparrow}$}\hspace*{-0.0cm} \framebox[0.125in] {$\uparrow \downarrow$}\hspace*{-0.0cm} \framebox[0.125in] {${\color{white}\uparrow} {\color{white}\uparrow}$}\hspace*{-0cm} \framebox[0.125in] {${\color{white}\uparrow} {\color{white}\uparrow}$}}$
and ${\tiny  \framebox[0.125in] {$\uparrow \downarrow$}\hspace*{-0.0cm} \framebox[0.125in] {$~~\uparrow{\color{white}\uparrow}$}\hspace*{-0.0cm} \framebox[0.125in] {$\uparrow \downarrow$} \hspace*{-0cm} \framebox[0.125in] {$~~\downarrow {\color{white}\uparrow}$}\hspace*{-0cm} \framebox[0.125in] {${\color{white}\uparrow} {\color{white}\uparrow}$}\hspace*{-0.0cm} \framebox[0.125in] {${\color{white}\uparrow} {\color{white}\uparrow}$}}$
with 12\% each for the $^1E$ state and
$ {\tiny  \framebox[0.125in] {$\uparrow \downarrow$}\hspace*{-0.0cm} \framebox[0.125in] {${\color{white}\uparrow}{\color{white}\uparrow}$} \hspace*{-0.0cm} \framebox[0.125in] {$\uparrow \downarrow$}\hspace*{-0.0cm} \framebox[0.125in] {$\uparrow \downarrow$}\hspace*{-0.0cm} \framebox[0.125in] {${\color{white}\uparrow} {\color{white}\uparrow}$}\hspace*{-0.0cm} \framebox[0.125in] {${\color{white}\uparrow} {\color{white}\uparrow}$}}$
with 20\% for the $^1A_1$ state. Refer to Table~\ref{tab:conf} for the notations.
The former states indicate single-excitations from the doubly occupied $a_1^C$ level,
while the latter state indicates a double-excitation from the $a_{1}^C$ level. These configurations have not been considered before in the
literature. Their inclusion in our work may have given rise to the discrepancy between our result and those obtained in Refs.~\cite{Bockstedte2018,Ma2020,ChoiPRB012} and it may also affect the intersystem crossing.

Furthermore, above the first-excited triplet $^3E$ state and the first-excited $^1A_1$ state, we find the triplet $^3A_1$ and $^3E$ states and
the singlet $^1A_2$ state. Due to the lack of experimental data beyond the four lowest states, we only briefly mention these higher-energy states.
Our higher-energy states differ from those in the literature~\cite{Bockstedte2018,ZyubinJPC07,Ma2010,ChoiPRB012,MazeNJP011,DohertyNJP011}.
As shown in Table~\ref{tab:conf}, the main contributions to these states originate from single excitations from the $a_1^C$, $e_x$, or $e_y$ orbital to beyond the dangling bond orbitals ($e_x^{\prime}$ and $e_y^{\prime}$) (Fig.~\ref{fig2}(b)). On the other hand, the previous many-body and molecular-model studies~\cite{Ma2010,ChoiPRB012,MazeNJP011,DohertyNJP011}
were mostly obtained considering only three or four dangling bond orbitals ($a_1^N$, $a_{1}^C$, $e_x$, and $e_y$). As discussed earlier, the
higher-energy states are more sensitive to the size of active space and cluster size than the four lowest states due to stronger electron
correlation. Note that the $^1E^{\prime}$ state predicted in the literature has not been experimentally observed~\cite{Kehayias2013}.

\subsection{Zero-field splitting}\label{ZeroField}

\begin{figure} 
\includegraphics[scale=0.55]{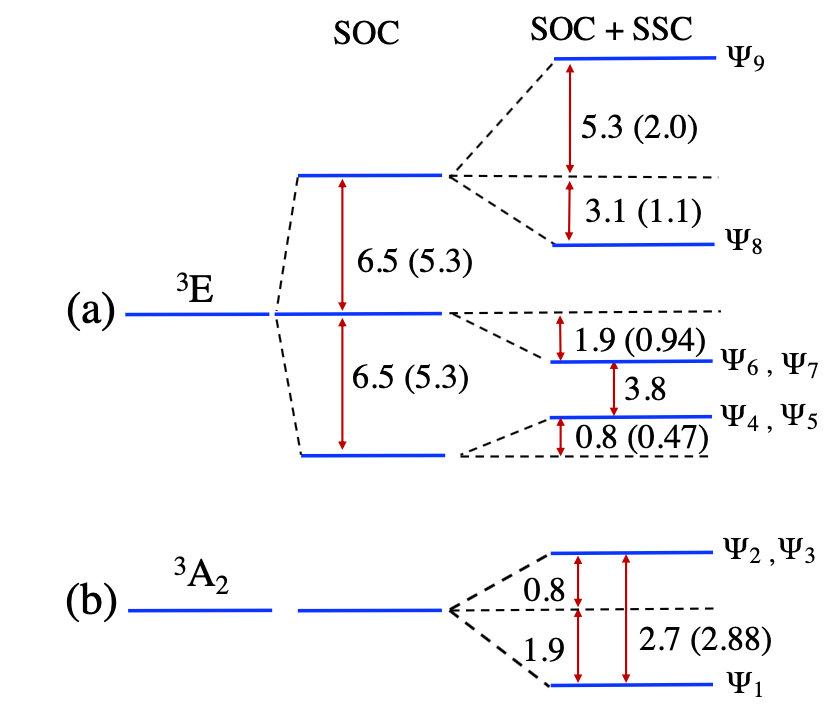}
\caption{Schematic diagram of our calculated energy level splitting of (a) the first-excited triplet $^3E$ state and (b) the ground
$^3A_2$ state due to SOC and SSC in units of GHz (for 70-atom cluster). The experimental values~\cite{BatalovPRL90} are shown inside parentheses.
States $\Psi_{1,...,9}$ are defined in Table~\ref{tab:eigenvectors}.} \label{fig4}
\end{figure}

\begin{table*}
\begin{ruledtabular}
\caption{Calculated SOC- and SSC-induced level splitting of the ground state ($^3A_2$) and the first-excited triplet ($^3E$) state from the quantum chemistry method for the 70-atom and 162-atom clusters in comparison to experiment. The level splitting values for the 162-atom cluster are shown in
the parentheses. All energies are expressed relative to the lowest SOC-included energy in each triplet state
($^3A_2$ or $^3E$). One exception is the experimental zero-field splitting of the $^3A_2$ state marked by ${\dagger}$ in which
only the difference is known. The eigenvectors including SOC, $\Psi_{1,...,9}$, are defined in Table~\ref{tab:eigenvectors}.\label{tab2}}
\footnotesize
\begin{tabular}{ll|ccc|cc}
 State &  & SOC (GHz) & SSC (GHz)  &  SOC+SSC      & SOC (GHz)    & SOC+SSC \\
       &  & (Theory)  & (Theory)   & (Theory, GHz) &  (Exp.)~\cite{BatalovPRL90} & (Exp., GHz)~\cite{BatalovPRL90} \\
     \hline
 $^3A_2$ & $\Psi_1$           & 0      &  $-$1.9   & $-$1.9   &    0   &  0$^{\dagger}$ \\
         & $\Psi_2$, $\Psi_3$ & 0      &  ~~0.8    & ~~0.8    &    0   &  2.88$^{\dagger}$ \\
         \hline
 $^3E$   & $\Psi_4$, $\Psi_5$ & 0      &  ~~0.8         & ~~0.8 (0.8)   &    0   &  0.47  \\
         & $\Psi_6$, $\Psi_7$ & 6.5 (8.1)    & $-$1.9    & ~~4.6 (6.2)    &   5.3  &  4.36   \\
         &  $\Psi_8$          & 13.0 (16.2)   & $-$3.1   & ~~9.9 (13.1)    &  10.6  &  9.52 \\
         &  $\Psi_9$          & 13.0 (16.2)    & ~~5.3   & ~~18.3 (21.5)   &  10.6  & 12.62  \\
\end{tabular}
\end{ruledtabular}
\end{table*}

\begin{table*}
\begin{ruledtabular}
\caption{Energy eigenvalues and eigenvectors corresponding to the ground- and first-excited triplet $^3A_2$ and $^3E$ states for the 70-atom and 162-atom clusters calculated using the quantum chemistry methods including SOC and SSC. The energies are relative to the lowest SOC-included energy of each triplet state ($^3A_2$ or $^3E$), as listed in Table~\ref{tab2} and shown in Fig.~\ref{fig4}. The energy values in the parentheses are for the 162-atom cluster. Here $\Psi_{1,T}$, $\Psi_{2,T}$, and $\Psi_{3,T}$ are our calculated eigenstates
(without SOC and SSC) listed in Table~\ref{tab:conf}.\label{tab:eigenvectors}}
\begin{tabular}{ l  l |l |l}
State  &      & Energy (GHz) & Total wave function \\ \hline
$^3A_2$ & $\Psi_{1}$ & $-$1.9 ($-1.9$)       & $\Psi_{1,T}$ $|S=1$, $M_z=0 \rangle $\\
        & $\Psi_{2}$ &    ~~0.8 (0.8)     & $\frac{1}{\sqrt{2}}$ $\Psi_{1,T}$ ($| S=1$, $M_z=1 \rangle$ $+$ $| S=1$, $M_z=-1 \rangle)$ \\
        & $\Psi_{3}$ &    ~~0.8 (0.8) & $\frac{1}{\sqrt{2}}$ $\Psi_{1,T}$ ($-$ $| S=1$, $M_z=1 \rangle$ $+$ $| S=1$, $M_z=-1 \rangle$) \\ \hline
$^3E$   & $\Psi_{4}$ &    ~~0.8 (0.8)     & $\frac{1}{\sqrt{2}}$($\Psi_{2,T} + i \Psi_{3,T}$) $| S=1$, $M_z=1 \rangle$) \\
        & $\Psi_{5}$ &    ~~0.8 (0.8)     & $\frac{1}{\sqrt{2}}$($\Psi_{2,T} - i \Psi_{3,T}$) $| S=1$, $M_z=-1 \rangle $) \\
        & $\Psi_{6}$ &    ~~4.6 (6.2)     & $\Psi_{2,T}$ $| S=1$, $M_z=0 \rangle$ \\
        & $\Psi_{7}$ &    ~~4.6 (6.2)     & $\Psi_{3,T}$ $| S=1$, $M_z=0 \rangle $ \\
        & $\Psi_{8}$ &    ~~9.9 (13.1)     & $\frac{1}{2} \Psi_{2,T}(| S=1, M_z=1 \rangle + | S=1, M_z=-1 \rangle )-i\frac{1}{2} \Psi_{3,T} (| S=1, M_z=1 \rangle - | S=1, M_z=-1 \rangle )$ \\
        & $\Psi_{9}$ &   ~~18.3 (21.5)     & $-\frac{1}{2}\Psi_{2,T}$ ($| S=1$, $M_z=1 \rangle $ $-$ $| S=1$, $M_z=-1 \rangle)+i\frac{1}{2}\Psi_{3,T} (| S=1, M_z=1 \rangle +| S=1, M_z=-1 \rangle)$\\
\end{tabular}
\end{ruledtabular}
\end{table*}

All of the spin-triplet states that we discussed earlier are split due to SOC and/or SSC. Note that SOC plays an important role in the
zero-field splitting only for the degenerate levels in this system because of the weak SOC. Since experimental data do not exist for higher-energy states, we present calculated zero-field splitting values of the ground $^3A_2$ state and the first-excited triplet $^3E$ state only.
Figure~\ref{fig4} and Table~\ref{tab2} show our calculated level splitting by SOC alone and by SOC in combination with SSC (SOC+SSC) for the $^3A_2$ and $^3E$ states, separately, compared to experimental data~\cite{BatalovPRL90}. Table~\ref{tab:eigenvectors} lists the corresponding eigenvectors $\Psi_{1,...,9}$ obtained from the quantum chemistry calculations including SOC and SSC. The SOC-induced level splitting is obtained for the 70-atom
and 162-atom clusters, while the SSC-induced splitting is obtained for the 70-atom cluster. Regarding the SOC+SSC induced splitting for the 162-atom
cluster, we use the SSC-induced splitting for the 70-atom cluster since the SSC-induced splitting does not depend much on cluster size. Let us now discuss the $^3A_2$ and $^3E$ states separately.

The SOC does not split the $^3A_2$ state to the first order and its splitting by higher-order SOC is negligible. However, we find that the SSC
splits the $^3A_2$ state into one lower non-degenerate level with $M_z=0$ and one higher doubly degenerate level with $M_z=\pm 1$ by $-$1.9~GHz and 0.8~GHz, respectively. (See the eigenvectors $\Psi_{1,2,3}$ in Table~\ref{tab:eigenvectors}.) Therefore, the energy separation between them is about 2.7~GHz, which is in excellent agreement with the experimental value of 2.88~GHz~\cite{BatalovPRL90} as well as a previous DFT calculation~\cite{Viktor014}.

On the other hand, the SOC splits the $^3E$ state into three (degenerate) groups, each of which has eigenvalues of the $z$ component of orbital angular momentum, $L_z$, of $\pm$0.46 ($\pm$0.53) for the 70-atom (162-atom) cluster. The separation of the levels is about 6.5~GHz for the 70-atom cluster (Fig.~\ref{fig4}) and about 8.1~GHz for the 162-atom cluster. Our calculated level splitting values show a weak cluster-size
dependence and they are somewhat larger than the experimental value of 5.3~GHz \cite{BatalovPRL90}. A possible reason for this is the dynamic Jahn-Teller effect \cite{Fu2009,Abtew2011} and the resulting quenching of SOC (i.e., Ham reduction factor \cite{Ham1968,Thierling2017,Goldman2017})
Note that our calculations are done for zero strain without electron-phonon coupling.
Quantum-chemistry calculations of electron-phonon coupling and the dynamic Jahn-Teller effect are worth investigating in the future.
In addition to the SOC-induced splitting, the SSC further shifts the lowest degenerate level upward by 0.8 GHz ($\Psi_{4}$, $\Psi_{5}$ in Table~\ref{tab:eigenvectors}) and moves the second degenerate level downward by 1.9 GHz ($\Psi_{6}$, $\Psi_{7}$). In this case, the degeneracy still holds. Interestingly, the amount of the downward level shift is almost twice that of the upward shift. The trend of the level-shift direction as well as the ratio between the downward and upward shift amount, are in good agreement with experiment~\cite{BatalovPRL90}, although our shifted values are off by a factor of 2 compared to experiment. We also find that the SSC splits the third doubly degenerate level into two separate levels ($\Psi_{8}$, $\Psi_{9}$): one level shifts downward by 3.1 GHz and the other moves upward by 5.3 GHz. Again, the trend of the
level shift agrees with experiment~\cite{BatalovPRL90}, although the calculated shift amount is greater than experiment by a factor of 2 or 3.
This overestimated SSC contribution may partially arise from our first-order perturbation treatment of SSC.

\section{Conclusion and Outlook}

We have developed a systematic numerical procedure to compute the electronic structure and magnetic properties of an NV$^-$ center defect in
diamond clusters, using the (multiconfigurational) quantum chemistry methods, where electron correlation is properly included. We found that the crucial constituent in the procedure is to identify and include extra unoccupied defect orbitals (beyond the four dangling bond orbitals) in the
active space. Our quantum chemistry calculations showed that the first-excited spin-triplet $^3E$ state is separated from the ground state ($^3A_2$)
by 1.93-2.14~eV, while the first-excited spin-singlet $^1E$ state is separated from the lower-energy $^1A_1$ state by 1.07-1.35~eV. In addition, we found that the $^3E$ state is separated from the $^1A_1$ state by 0.52-0.54~eV. Our calculated triplet-triplet, singlet-singlet, and triplet-singlet excitation energies as well as the ordering of the triplet and singlet states are in good agreement with experiment. 
We found additional configurations which significantly contribute to the $^1E$ and $^1A_1$ states, which have not been considered before.
Furthermore, SOC and SSC were included in our many-body wave functions, finding that the SSC splits the $^3A_2$ state by 2.7~GHz and that a combination of the SOC and SSC splits
the $^3E$ state into two degenerate levels and two non-degenerate levels. The SSC-induced splitting of the $^3A_2$ state and the SOC-induced splitting of the $^3E$ state in good agreement with experiment. When both SOC and SSC are included in the $^3E$ state, the calculated trend of the level splitting agree well with experiment and the splitting amount is mostly deviated from experiment by a factor of two.


The numerical procedure that we developed in this work can be applied to other deep defects in wide band-gap semiconducting materials such as
group-IV defects and transition-metal defects in diamond or silicon carbide, or rare-earth defects in silicon or complex oxides, as long as a sufficient number of defect-localized orbitals is judiciously chosen for the active space while retaining the defect symmetries and orbital
degeneracy as accurately as possible. This procedure may also be extended to obtain radiative transition rates between the states and can be applied to deep defects with external perturbations such as electric fields and strains. Therefore, our findings open a new avenue to be able to screen other defects desirable for specific applications beyond to accurately predict the properties of their excited states.

\acknowledgements{
The authors were supported by the National Science Foundation under a collaborative grant (Grant numbers DMR-1737921 for S.E.E. and K.P. and DMR-1738076 for P.D.). We are grateful to Adam Gali, Kai-Mei Fu, Marcus Doherty, and Audrius Alkauskas for useful comments on the manuscript.
The computational support was provided by the
Virginia Tech Advanced Research Computing and the Extreme Science and Engineering Discovery Environment (XSEDE) under Project number DMR060009N (K.P.) and Project number PHY180014 (P.D.), which are supported by the National Science Foundation Grant number ACI-1548562.}

\appendix
\section{Procedure of identifying active orbitals and performing CASSCF(6,6)}\label{appendix1}

\begin{figure}[hbt!]
\centering
\includegraphics[scale=0.5]{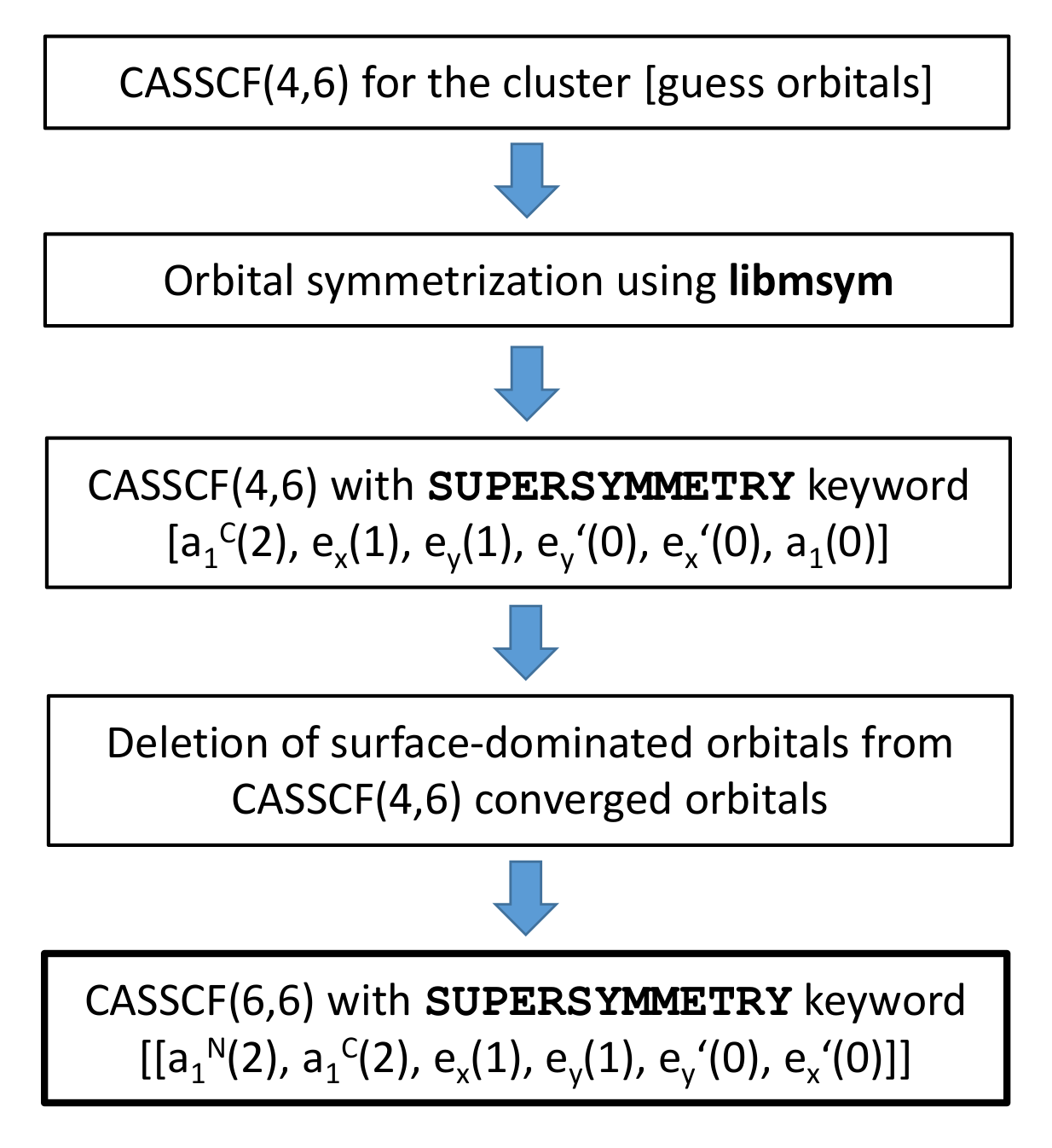}
\caption{Schematic diagram of our practical procedure to identify two extra unoccupied defect orbitals and to preform the CASSCF(6,6)
calculations of an NV$^{-}$ center defect in the hydrogen-passivated 70-atom and 162-atom diamond clusters, using {\tt OpenMolcas}.
Here initial orbitals in the active space are listed within brackets, where nominal occupation numbers for the spin-triplet ground state
are shown inside parentheses. The nominal occupation numbers differ from the actual occupation numbers. The orbitals inside double brackets
are final converged orbitals. {\tt libmsysm} is an orbital-symmetrization program~\cite{Johnson017} and the function of
{\tt SUPERSYMMETRY} keyword is defined in the text of the Appendix.}
\label{fig:CASSCF}
\end{figure}

In order to identify extra unoccupied defect orbitals beyond the four dangling bond orbitals as discussed in Section 4, we carry out the following systematic procedure for the 70-atom and 162-atom clusters with the total spin $S=1$. Figure~\ref{fig:CASSCF} summarizes the CASSCF procedure using {\tt OpenMolcas}. Note that extra unoccupied defect orbitals cannot be found from the CASSCF(6,4) calculation. The doubly occupied $a_1^N$
orbital is known to have a lower energy than the doubly occupied $a_1^C$ orbital and the former is be buried in the bulk valence band. Therefore,
excluding the $a_1^N$ orbital, we envision a CASSCF(4,6) calculation where six active orbitals consist of three dangling bond orbitals
($a_1^C$, $e_x$, $e_y$), two unoccupied defect orbitals with $E$ IRRep, and one unoccupied defect orbital with $A_1$ IRRep. Keeping this in mind, we first perform a CASSCF(4,6) calculation (with state average over six roots) using four active electrons and initial six active orbitals guessed by
{\tt OpenMolcas}. Then converged orbitals from the CASSCF(4,6) calculation are fully symmetrized with $C_{3v}$ symmetry, using the {\tt libmsym} program~\cite{Johnson017} that is interfaced with {\tt OpenMolcas}. The {\tt libmsym} program can deal with higher point-group symmetries than twofold symmetry. Now each molecular orbital in the inactive, active and virtual spaces has its own pure IRRep symmetry. Among these symmetrized orbitals, we identify two extra unoccupied orbitals localized near the defect with $e_x$ and $e_y$ symmetries, as well as one unoccupied defect orbital with $a_1$ symmetry. In order to distinguish these extra orbitals with $e_x$ and $e_y$ symmetries from the singly occupied dangling bond orbitals
($e_x$ and $e_y$), the former orbitals are referred to as $e_x^{\prime}$ and $e_y^{\prime}$ orbitals. Now using these extra three unoccupied defect
orbitals as well as the three dangling bond orbitals as initial six active orbitals, we carry out another CASSCF(4,6) calculation with restricted orbital rotations throughout iterations, in other words, orbital rotations (or optimization) are allowed only among the orbitals belonging to the same IRRep. This restriction can be achieved using {\tt SUPERSYMMETRY} keyword in {\tt OpenMolcas} code. The steps of libmsym and {\tt SUPERSYMMETRY} are crucial to maintain purely-symmetric orbitals throughout the self-consistent calculations and more importantly to retain the perfect degeneracy of the converged CASSCF energy eigenvalues (the accuracy of $\sim 10$~neV) belonging to the IRRep $E$. Such high accuracy is required for an accurate calculation of zero-field splitting induced by SOC. After the second CASSCF(4,6) calculation, the two unoccupied defect orbitals, $e_x^{\prime}$ and $e_y^{\prime}$, remain in the active space.

In our molecular cluster models for an NV$^{-}$ center, the hydrogen-passivated surface is artificial since it does not exist in a diamond lattice. Therefore, orbitals localized at the surface are not associated with the defect in a diamond lattice. In order to reduce an effect of such surface-dominated orbitals on the orbital optimization, we remove several tens of surface-dominated orbitals near the active space from
the converged orbitals in the second CASSCF(4,6) calculation. More surface orbitals are removed for a larger cluster. After this step, we now
carry out a CASSCF(6,6) calculation with {\tt SUPERSYMMETRY} keyword using the identified $e_x^{\prime}$ and $e_y^{\prime}$ orbitals (from the
CASSCF(4,6) calculation) as well as the four dangling bond orbitals as initial active orbitals.
We check that the energy levels (root energies) obtained from the CASSCF(6,6) calculation do not change as the number of removed surface
orbitals varies, as long as enough number of surface orbitals are removed near the active space.

The similar procedure to Fig.~\ref{fig:CASSCF} is carried out for the total spin $S=0$ with state average over six roots for both 70-atom and 162-atom
clusters. Then we perform another CASSCF(6,6) calculation with state average over four roots, using the converged CASSCF(6,6) orbitals,
in order to retain the perfect degeneracy of the CASSCF energy eigenvalues in the $E$ IRRep and the localization of the active orbitals.
We emphasize that the orbital symmetrization is more important for the spin-singlet states than for the spin-triplet states.

\bibliography{molcas,orca,NV}
\bibliographystyle{unsrt}

\end{document}